\documentstyle[eqsecnum,prd,aps,epsfig]{revtex}
\newcommand{\dfr}{d\raise0.3ex\hbox{\kern-0.6ex\char"013 }} 
\newcommand{\ddfr}{{ }^{-}\!\!\!\!\!\dfr} 
\begin{document}
%
%
\def\ov{\over}
\def\l{\left}
\def\r{\right}
\def\be{\begin{equation}}
\def\ee{\end{equation}}
\draft
\title{Dirac Equation in Scale Relativity}
\author{Marie-No\"elle C\'el\'erier and Laurent Nottale}
\address{LUTH, CNRS, Observatoire de Paris-Meudon, 
5 place Jules Janssen, 92195 Meudon Cedex, France}

\date{\today}
\maketitle

\begin{abstract}
The theory of scale relativity provides a new insight into the origin 
of fundamental laws in physics. Its application to microphysics allows 
to recover quantum mechanics as mechanics on a non-differentiable 
(fractal) space-time. The Schr\"odinger and Klein-Gordon equations 
have already been demonstrated as geodesic equations in this framework. 
We propose here a new development of the intrinsic properties of this 
theory to obtain, using the mathematical tool of Hamilton's bi-quaternions, 
a derivation of the Dirac equation, which, in standard physics, is merely 
postulated. The bi-quaternionic nature of the Dirac spinor is obtained 
by adding to the differential (proper) time symmetry breaking, which yields 
the complex form of the wave-function in the Schr\"odinger and Klein-Gordon 
equations, the breaking of further symmetries, namely, the differential 
coordinate symmetry ($dx^{\mu} \leftrightarrow - dx^{\mu}$) and 
the parity and time reversal symmetries.
\end{abstract}

\pacs{PACS numbers: 03.65.Pm, 12.90.+b}

\section{Introduction}
\label{s:intro}

The theory of scale relativity consists of generalizing to 
scale transformations the principle of relativity, which has been applied 
by Einstein to motion laws. It is based on the giving up of the assumption 
of space-time coordinate differentiability, which is usually retained as an 
implicit hypothesis in current physics. Even though this hypothesis can be 
considered as roughly valid in the classical domain, it is clearly broken 
by the quantum mechanical behavior. It has indeed been pointed out by 
Feynman (see, e.g., Ref. \cite{FH65}) that the typical paths of quantum 
mechanics are continuous but non-differentiable. \\

In the present paper, after a reminder about the Schr\"odinger and 
Klein-Gordon equations, we apply the scale-relativistic approach to the 
issue of the understanding of the nature of bi-spinors and of the Dirac 
equation in a space-time representation. A step in this direction has 
been made by the work of Gaveau et al. \cite{GJ84} who have generalized 
Nelson's stochastic mechanics \cite{EN66} to the relativistic case in (1+1) 
dimensions. However, an analytic continuation was still needed to obtain 
the Dirac equation in such a framework, and, furthermore, stochastic 
mechanics is now known to be in contradiction with standard quantum 
mechanics \cite{WL93}.\\

Basing himself on the concept of fractal space-time as a 
geometric analog of quantum mechanics, Ord \cite{GO83,NS84} has developed a 
generalized version of the Feynman chessboard model which allows 
to recover the Dirac equation in (3+1) dimensions, without 
analytical continuation \cite{GO92,MO92,OM93,GO96}. 
We develop in the present paper 
an approach involving also a fractal space-time, but in a different 
way: namely, the fractality of space-time is derived from the giving up of its 
differentiability, it is constrained by the principle of scale relativity, and 
the Dirac equation is derived as an integral of the geodesic equation. 
This is not a stochastic approach (the use of statistically defined quantities 
is derived and we do not use Fokker-Planck equations), so that it does 
not come under the contradictions encountered by stochastic mechanics. \\

One of the fundamental reasons for jumping to a scale-relativistic description 
of nature is to increase the generality of the first principles 
retained to construct the theory. Since more than three centuries, i.e., 
since Newton and Leibnitz, the founders of the integro-differential 
calculus, physics relies on the assumption that coordinates, and more 
generally all the physical equations, are a priori differentiable. 
Indeed, one of the most powerful avenue for reaching a genuine 
understanding of the laws of nature has always been to construct them by 
attempting to lessen the number of a priori assumptions retained, a 
principle known as Occam's razor. From this point of view, the laws and 
structures of nature are simply the most general laws and structures that 
are physically possible. It culminates nowadays in Einstein's principle 
of ``general'' relativity and his description of gravitation from its 
analogy with the manifestation of a Riemannian geometry of space-time, 
obtained by giving up the Euclidean flatness hypothesis. \\

However, in the light of the above remark, Einstein's principle of 
relativity is not yet fully general, since it applies to coordinate 
transformations that are continuous and, at least two times, differentiable. 
The aim of the theory of scale relativity is to bring to light laws and 
structures that would be the manifestation of still more general 
transformations, namely, continuous ones, either differentiable or not. The 
standard ``general relativistic'' theory will be automatically recovered as 
the special differentiable case, while new effects would be expected from the 
non-differentiable part. \\

Giving up the assumption of differentiability has important physical 
consequences: one can show \cite{LN93,LN94A} that spaces of topological 
dimension $D_T$, which are continuous but non-differentiable, are 
characterized by a $D_T$ measure which becomes explicitly dependent on 
the resolution (i.e., the observation scale) $\epsilon$ at which it is 
considered and tends to infinity 
when the resolution interval $\epsilon$ tends to zero. Therefore, 
a non-differentiable space-time continuum is necessarily fractal, in the 
general meaning given to this word by Mandelbrot \cite{BM82}, who thus 
means, not only self-similar, but explicitly scale-dependent in the more 
general way. This statement 
naturally leads to the proposal of a geometric tool adapted to construct 
a theory based on such premises, namely, a fractal space-time (see, e.g., 
Refs. \cite{GO83,NS84,LN93,LN89,EN92,BC00}). 
This allows to include the resolutions in the definition 
of the state of the reference system and to require scale covariance of 
the equations of physics under scale transformations. A main consequence 
is that the geodesics of such a non-differentiable space-time are 
themselves fractal and in infinite number between any two points. This 
strongly suggests that the description could borrow some tools from the 
mathematical formalism of statistics, i.e., will 
provide a probabilistic interpretation, therefore naturally extending the 
realm of quantum behaviour to a larger spectrum of available scales. \\

The domains of application are typically the asymptotic scale ranges of 
physics: (i) {\it Microphysics}: small length-scales and time-scales. (ii) 
{\it Cosmology}: large length-scales. In this framework, new solutions can 
be brought to the problems of the vacuum energy density, of Mach's 
principle and of Dirac's large number hypothesis, leading to a theoretical 
prediction of the value of the cosmological constant \cite{LN96A}. (iii) 
{\it Complex chaotic systems}: long time-scales. It has been suggested 
that a general theory of structure formation could take a quantum-like 
form, but with an interpretation differing from that of standard quantum 
mechanics \cite{LN93,LN92}. Several applications to the problem of the 
formation and evolution of gravitational structures have been completed 
in the recent years: solar system \cite{LN93,NSG97,LN98A}, extra-solar 
planetary systems \cite{LN96B}, planets around pulsars \cite{LN98B}, 
satellites of giant planets \cite{HSG98}, binary stars \cite{NLS97}, binary 
galaxies \cite{NT98}, large-scale distribution of galaxies \cite{LN96A}. 
Moreover, an additional scalar field appears naturally in such a construction 
(as a manifestation of the fractality of space-time, in analogy with the 
Newton field arising from its curvature), which could explain the various 
effects usually attributed to unseen, ``dark matter'' \cite{LN01}. 
It has also been suggested 
that this approach could be applied to the 
morphogenesis of a wide class of hierarchical structures, even outside the 
strict field of physics: for instance, in biology, structures are ``density 
waves'' rather than solid objects and could therefore be described in terms 
of probability amplitudes, being themselves solutions of a generalized 
Schr\"odinger equation \cite{LN97A}. \\

The theory of scale relativity can also be considered, from the point of 
view of its construction, as a completion of the standard laws of classical 
physics (motion in space, displacement in space-time) by new scale laws, 
in which the space-time resolutions are used as intrinsic variables, playing 
for scale transformations the same role as velocities for motion 
transformations. 
We hope such a stage of the theory to be only provisional, and the motion 
and scale laws to be treated on the same footing in the complete theory. 
But before reaching such a goal, various possible combinations of scale 
and motion laws, leading to a number of sub-sets of the theory, have to 
be developed. \\

We can first distinguish: (1) {\it Pure scale laws}: description of the 
internal structures of a non-differentiable space-time at a given 
point. (2) {\it Induced effects of scale laws on the equations of 
motion}: generation of the quantum mechanichal behavior as motion in a 
non-differentiable space/space-time. (3) {\it Scale-motion coupling}: 
effects of dilations induced by displacements, that have been tentatively 
interpreted as gauge fields, allowing to suggest the existence of new 
relations between the masses and the charges of elementary particles 
\cite{LN96A,LN94B}. \\

Several levels of the description of pure scale laws (1) can also 
be distinguished, which therefore induce new sub-sets inside each of the three 
above cases. These levels can actually be considered as reproducing 
the historical development of the theory of motion: (a) {\it Galilean 
scale relativity}: laws of dilation exhibiting the structure of a Galileo 
group, i.e., a fractal power law with a constant fractal dimension. 
(b) {\it Special scale relativity}: generalization of the laws of dilation 
to a Lorentzian form. The fractal dimension becomes a variable and is 
assimilated to a fifth dimension. A finite impassable length/time-scale, 
invariant under dilations, shows up. It owns all the physical properties 
of a zero (at small scales) or of an infinite (at large scales) - an 
infinite energy-momentum would be needed to reach it - and it plays for 
scale laws the same role as played by light velocity for motion. 
(c) {\it Scale dynamics}: while the first two above cases correspond to 
``scale freedom", one can also consider the distorsion from strict 
self-similarity which would proceed from the effect of a ``scale force" or 
``scale field" \cite{LN97A,LN97B}. (d) {\it General scale relativity}: in 
analogy with the gravitation field being assimilated to the effect of 
the space-time geometry in Einstein's general motion relativity, a 
future description of the scale field of (c) could 
be done in terms of a Riemannian geometry of the five-dimensional scale 
space, therefore assuming differentiability of scale transformations.
(e) {\it Quantum scale relativity}: the scale transformations are still 
assumed to be continuous but no more differentiable (as it has been 
previously done for space-time). One is therefore confronted, for scale 
laws, to the same conditions that lead to a quantum-like behaviour in a 
fractal space-time, and may conjecture that quantum scale laws could 
emerge at this (to be developed) level of the theory. \\

A forth spliting of the different levels and sub-levels described above 
is to be considered, depending on the status of the motion laws: 
non-relativistic, special-relativistic, general-relativistic. \\

In the present work, we focus our attention on the microphysical 
scale of motion in a non-differentiable space-time, in the framework of 
Galilean scale relativity (i2a). Therefore, we recover, from the first 
principles of this theory, the three main evolution equations of standard 
quantum physics as geodesic equations on a fractal space/space-time: \\
- the Schr\"odinger equation \\
- the (free) Klein-Gordon equation \\
- the (free) Dirac equation \\
which are here demonstrated, while they are merely postulated in standard 
quantum mechanics (the compatibility between the Dirac equation and the 
scale-relativistic formalism has already been studied by Pissondes 
\cite{JP99A}). \\

A reminder of the derivation of the Schr\"odinger equation is given in 
Sec.~\ref{s:schro}, where we actualize the former works dealing with this 
issue, proposing a more accurate and profound interpretation of the nature 
of the transition from the non-differentiable (small scales) to the 
differentiable (large scales) domain, and carefully justifying the 
different key choices made at each main step of the reasoning. An 
analogous reminder and updating is proposed in Sec.~\ref{s:ckgeq}, for 
the free Klein-Gordon equation exhibiting a complex (number) form. Then 
we derive, in Sec.~\ref{s:kgeq}, the bi-quaternionic form of 
the same Klein-Gordon equation, from which the Dirac equation naturally 
proceeds, as shown in Sec.~\ref{s:dieq}. The two appendices are devoted to the 
presentation and justification of the mathematical tools used in this 
framework. \\

The reader already familiar with the scale-relativistic formalism, and mainly 
interested in the broad new results presented here, can limit his attention 
to Secs.~\ref{s:kgeq} and ~\ref{s:dieq}, after a detour via 
Sec.~\ref{ss:fracsp}, where a new symbolism is introduced.

\section{The Schr\"odinger equation revisited}
\label{s:schro}

In previous works, the Schr\"odinger equation has already been derived, in the 
framework of scale relativity, as a geodesic equation in a fractal three-space 
(see, e.g., Refs. \cite{LN93,LN96A,LN97A}). We give, in the present section, a 
reminder of the main steps of the reasoning leading to its formulation, 
therefore making the reader familiar with the structure and tools of the 
theory. We also update and give a more precise meaning to some of 
the most subtle issues we are dealing with.

\subsection{Scale invariance and Galilean scale relativity}
\label{ss:galsr}

A power law scale dependence, frequently encountered in natural 
systems, is described geometrically in terms of fractals \cite{BM82,BM75}, 
and algebrically in terms of the renormalization group \cite{KW75,KW79}. 
As we show below, such simple scale-invariant laws can be identified 
with the ``Galilean" version of scale-relativistic laws. \\

Consider a non-differentiable (fractal) curvilinear coordinate 
${\cal L}(x,\epsilon)$, 
that depends on some space-time variables $x$ and on the resolution 
$\epsilon$. Such a coordinate generalizes to non-differentiable and fractal 
space-times the concept of curvilinear coordinates introduced for curved 
Riemannian space-times in Einstein's general relativity \cite{LN93}. Rather 
than considering only the strict non-differentiable mathematical object 
${\cal L}(x)$, we are interested in its various resolutions $\epsilon$. 
Such a point of view is particularly well-adapted to applications in physics, 
as any real measurement is always performed at a finite resolution. In this 
framework, ${\cal L}(x)$ becomes the limit when $\epsilon \rightarrow 0$ 
of the family of functions ${\cal L}(x,\epsilon)$. But while ${\cal L}(x,0)$ 
is non-differentiable, ${\cal L}(x,\epsilon)$, which we call a fractal 
function, is differentiable for all $\epsilon \neq 0$ (see Refs. 
\cite{BC00,JC01} 
for a precise mathematical definition of a fractal function). The physics 
of a given process will therefore be completely described if we succeed in 
knowing ${\cal L}(x,\epsilon)$ for all values of $\epsilon$. 
${\cal L}(x,\epsilon)$, being differentiable when $\epsilon \neq 0$, can 
be the solution to differential equations involving the derivatives of 
${\cal L}$ with respect to both $x$ and $\epsilon$. \\

Let us now apply an infinitesimal dilation $\epsilon \rightarrow 
\epsilon '=\epsilon(1+d\rho)$ to the resolution. Being, at this stage, 
interested in pure scale laws, we omit the $x$ dependence in order to 
simplify the notation and obtain, at first order,
\be
{\cal L}(\epsilon ')={\cal L}(\epsilon +\epsilon d\rho)={\cal L}(\epsilon)+
{{\partial {\cal L}(\epsilon)}\over {\partial \epsilon}} \epsilon d\rho=
(1+\tilde{D} d\rho){\cal L}(\epsilon),
\label{eq.1}
\ee
where $\tilde{D}$ is, by definition, the dilation operator. The identification 
of the two last members of this equation yields 
\be
\tilde{D}=\epsilon{\partial \over {\partial \epsilon}}={\partial \over 
{\partial 
\ln \epsilon}} \; .
\label{eq.2}
\ee

This well-known form of the infinitesimal dilation operator shows that the 
``natural'' variable for the resolution is $\ln \epsilon$, and that the 
expected new differential equations will indeed involve quantities as 
$\partial {\cal L}(x,\epsilon)/\partial \ln \epsilon$. The equations 
describing scale dependence are known to be the renormalization group 
equations, as first proposed by Wilson \cite{KW75,KW79}. \\

We intend to limit the present approach to the consideration of the simplest 
form that can be exhibited by this class of equations, leaving to future 
works the study of more complete cases. We therefore interpret the 
simplest renormalization group-like equation implying an essential physical 
quantity like ${\cal L}$ as stating that the variation of ${\cal L}$ under 
an infinitesimal scale transformation $d\ln\epsilon$ depends only on 
${\cal L}$ itself; namely, ${\cal L}$ determines the whole physical behavior, 
including the behavior under scale transformations. We thus write
\be
{\partial {\cal L}(x,\epsilon)\over {\partial \ln \epsilon}}=\beta(\cal L).
\label{eq.3}
\ee

Still interested in the simplest form for such an equation, we expand 
$\beta(\cal L)$ in powers of $\cal L$. This can always be done since 
$\cal L$ may be renormalized, dividing it by its largest value, 
in such a way that the new variable  $\cal L$ remains $\ll 1$ in its 
variation domain. We obtain, to the first order, the linear equation
\be
{\partial {\cal L}(x,\epsilon)\over {\partial \ln \epsilon}}=a+b\cal L \; ,
\label{eq.4}
\ee
of which the solution is 
\be
{\cal L}(x,\epsilon) = {\cal L}_0(x)\; \left[1+\zeta (x)(\frac{\lambda}
{\epsilon})^{-b}\right],
\label{eq.5}
\ee
where $\lambda ^{-b} \zeta(x)$ is an integration constant and 
${\cal L}_0=-a/b$. \\

For $b<0$, Eq.~(\ref{eq.5}) gives a fractal scale-invariant behavior at 
small scales ($\epsilon \ll \lambda$), with a fractal dimension 
$D=D_T-b$. For a curvilinear coordinate ${\cal L}$, the topological dimension 
$D_T$ is equal to $1$. The scale dimension, $\delta = D - D_T$, is defined, 
following Mandelbrot \cite{BM82,BM75}, as 
\be
\delta = \frac{d \ln {\cal L}}{d \ln ({\lambda/ \epsilon})} \; .
\label{eq.6}
\ee

When $\delta$ is constant, one obtains an asymptotic power law resolution 
dependence 
\be
{\cal L}(x,\epsilon) = {\cal L}_0(x) \left(\lambda\over\epsilon\right)^\delta.
\label{eq.7}
\ee

We have stressed in the introduction the fact that the new scale laws to 
be constructed were expected to come under the principle of relativity 
extended to scale transformations of the resolutions $\epsilon$. Let 
us now check that this is indeed the case for such a simple self-similar 
scaling law. The Galilean structure of the group of scale transformations 
which correspond to this law can be verified, in a straightforward manner, 
from the fact that the involved quantities
transform, under a scale transformation $\epsilon \rightarrow 
\epsilon '$, as
\be
\ln \frac{{\cal L} (\epsilon ')}{{\cal L}_0} = 
\ln \frac{{\cal L} (\epsilon)}{{\cal L}_0} +  
\delta (\epsilon)\ln \frac{\epsilon}{\epsilon '} \; ,
\label{eq.8}
\ee
\be
\delta(\epsilon ') = \delta(\epsilon).
\label{eq.9}
\ee

These transformations have exactly the structure of the Galileo group, as 
confirmed by the dilation composition law, 
$\epsilon \rightarrow \epsilon ' \rightarrow \epsilon ''$, which writes
\be
\ln {\epsilon ''\over \epsilon} = \ln {\epsilon '\over \epsilon} + 
\ln {\epsilon ''\over \epsilon '} \; . 
\label{eq.10}
\ee

It is worth noting that Eq.~(\ref{eq.5}) gives also a transition from a 
fractal to a non-fractal behavior at scales larger than some transition scale 
$\lambda$. \\

The solutions corresponding to the case $b>0$ yield scale 
dependence at large scales, broken to scale independence below $\lambda$. 
While $b<0$ is characteristic of the microphysical situation (it yields 
both quantum phenomena and the quantum-classical transition), the case $b>0$ 
is encountered in the astrophysical and cosmological domains (see, e.g., 
Ref. \cite{LN97A}). \\

We want to stress that we have retained here the simplest case of a first 
order limited expansion of $\beta({\cal L})$ in powers of ${\cal L}$. Going to 
higher order in Eq.~(\ref{eq.4}) would produce an increasing number 
of transition scales: two for a second order limited expansion, three for 
a third order, and so on.

\subsection{Lagrangian approach to scale laws}
\label{ss:lagrang}

We are naturally led, in the scale-relativistic approach, to reverse the 
definition and meaning of the variables. The scale dimension $\delta$ 
becomes a fundamental variable. It plays, for scale laws, the same role as 
played by (proper) time in motion laws. It remains constant in only very 
peculiar situations, namely in the case of scale invariance, corresponding to 
``scale freedom'', which is the (a) case here studied. When going to a more 
general description of pure scale laws (b-c-d-e), one is led to consider 
a scale dimension $\delta(\epsilon)$, which is no more a constant 
\cite{LN93}. \\

The resolution, $\epsilon$, can therefore be defined as a derived quantity in 
terms of the fractal coordinate $\cal L$ and of the scale dimension $\delta$ 
\be
{\bar{V}} = \ln ({\lambda \over \epsilon}) = \frac{d \ln {\cal L}}{d \delta} 
\; .
\label{eq.11}
\ee

Our identification of the standard fractal behavior with Galilean scale laws 
can now be fully justified. We assume that, as in the case of motion laws, 
the scale laws can be constructed from a Lagrangian approach. A scale Lagrange 
function ${\bar{L}}(\ln {\cal L}, {\bar{V}}, \delta)$ is introduced, from 
which a scale action is constructed
\be
{\bar{S}} = \int_{\delta_1}^{\delta_2} {\bar{L}}(\ln {\cal L}, {\bar{V}}, 
\delta) d \delta.
\label{eq.12}
\ee

The application of the action principle yields a scale Euler-Lagrange 
equation that writes
\be
\frac{d} {d \delta} \frac{\partial {\bar{L}}} {\partial{\bar{V}}} = 
\frac{\partial {\bar{L}}} {\partial \ln {\cal L}} \; .
\label{eq.13}
\ee

The simplest possible form for the Lagrange function is a quadratic 
dependence on the ``scale velocity'', $\bar V$, (i.e., $ 
{\bar{L}}~\propto~{\bar{V}}^2$) and the absence of any ``scale force'' 
(i.e., $\partial{\bar{L}} / \partial \ln {\cal L} = 0$), which is the 
equivalent for scale of what inertia is for motion. Note that this form of 
the Lagrange function becomes fully justified, as in the case of motion laws, 
when one jumps to special scale relativity, then back to 
the Galilean limit \cite{LN97A}. The Euler-Lagrange equation becomes in this 
case
\be
\frac{d{\bar{V}}}{d \delta} = 0 \Rightarrow {\bar{V}} = const.
\label{eq.14}
\ee

The constancy of ${\bar{V}} = \ln (\lambda/\epsilon)$ means that it is 
independent of the ``scale time'' $\delta$. Equation (\ref{eq.11}) can 
therefore be integrated to give the usual power law behavior, 
${\cal L} = {\cal L}_0 (\lambda/\epsilon)^\delta$. This reversed viewpoint 
has several advantages which allow a full implementation of the principle 
of scale relativity: \\
(i) The scale dimension $\delta$ is given its actual status of ``scale time" 
and the logarithm of the resolution, ${\bar{V}}$, its status of 
``scale velocity" (see Eq.~(\ref{eq.11})). This is in accordance with its 
scale-relativistic definition, in which  it characterizes the ``state of 
scale'' of the reference system, in the same way as the velocity 
$v = dx/dt$ characterizes its state of motion. \\
(ii) This leaves open the possibility of generalizing our formalism to the 
case of four independent space-time resolutions, ${\bar{V}}^{\mu} = 
\ln (\lambda^{\mu} / \epsilon^{\mu}) = d \ln {\cal L}^{\mu}/ d \delta$.\\
(iii) Scale laws more general than the simplest self-similar ones can be 
derived from more general scale Lagrangians \cite{LN97A}.

\subsection{Transition from non-differentiability (small scales) to  
differentiability (large scales)}
\label{ss:fracsp}

The aim of the present subsection is to identify the consequences of the 
giving up of the coordinate differentiability. We develop here the 
formalism relevant for the derivation 
of the non-relativistic motion Schr\"odinger equation, valid in a fractal 
three-space with the time $t$ as a curvilinear parameter. This formalism will 
be easily generalized, in Sec.~\ref{s:ckgeq}, to 
relativistic motion involving a four-dimensional space-time with the 
proper time $s$ as a curvilinear parameter. \\

Strictly, the non-differentiability of the coordinates implies that the 
velocity
\be
V = {dX\over dt}= \lim_{dt \rightarrow 0} \frac{X(t+dt) - X(t)}{dt}
\label{eq.15}
\ee
is undefined. This means that, when $dt$ tends to zero,  either the ratio 
$dX/dt$ tends to infinity, or it fluctuates without reaching any limit. 
However, as recalled in the introduction, continuity and non-differentiability 
imply an explicit scale dependence of the various physical quantities, 
and therefore of the velocity, $V$. We thus replace the differential, $dt$, by 
a scale variable, $\delta t$, and consider $V(t,\delta t)$ as an explicit 
fractal function of this variable. In the case of a constant fractal 
dimension $D$, the resolution in $X$, $\epsilon(\delta t)$, corresponds to 
the resolution in $t$, $\delta t$, according to the relation \cite{LN93}
\be
\epsilon_x \approx \delta t ^{1/D}.
\label{eq.16}
\ee

The advantage of this method is that, for any given value of the resolution, 
$\delta t$, differentiability in $t$ is recovered, which allows to use 
the differential calculus, even when dealing with non-differentiability. 
However, one should be cautious about the fact that the physical and the 
mathematical descriptions are not always coincident. Indeed, once $\delta t$ 
is given, one can write mathematical differential equations involving 
$\partial / \partial t$, make $\partial t \rightarrow 0$, then solve for 
it and determine $V(t, \delta t)$. Actually, this is a purely mathematical 
description with no physical counterpart, since, for the system under 
study, the very consideration of an interval $dt < \delta t$, as occuring 
in an actual measurement, changes the 
function $V$ (such a behavior, described by Heisenberg's uncertainty 
relations, is experimentally verified for any quantum system). However, 
there is a peculiar subspace of description where the physics and the 
mathematics coincide, namely, when making the particular choice 
$dt = \delta t$. We work, for the Schr\"odinger and complex Klein-Gordon 
equations, with such an identification of the time 
differential and of the new time resolution variable, but are led to 
give it up in  Secs.~\ref{s:kgeq} and ~\ref{s:dieq}. \\

The scale dependence of the velocity suggests that we complete the 
standard equations of physics by new differential equations of scale. We 
therefore apply to the velocity and to the differential element, now 
interpreted as a resolution, the reasoning applied to the fractal function 
$\cal L$ in Sec.~\ref{ss:galsr}. Writing the simplest possible equation for 
the variation of the velocity $V(t, dt)$ in terms of the new scale variable 
$dt$, as a first order renormalization group-like differential equation 
of the form of Eq.~(\ref{eq.3}); then, Taylor-expanding it, as in 
Eq.~(\ref{eq.4}), using the fact that $V<1$ (in motion-relativistic units 
$c=1$), we obtain the solution
\be
V = v + w = v \; \left[1 + \zeta \left(\frac{\tau}{dt}\right)^{1-1/D}\right],
\label{eq.17}
\ee
where we have set $b=1/D-1$. Here, $v$ is the large-scale part of the 
velocity, $\overline{LS}<V>$, (see below the definition of the large-scale 
part operator $\overline{LS}$), $w$ is an explicitly scale-dependent fractal 
fluctuation and $\tau$ and $\zeta$ are chosen such that 
$\overline{LS}<\zeta> = 0$ and $\overline{LS}<\zeta^2> = 1$. \\

We recognize here the combination of a typical fractal behavior, with a 
fractal dimension $D$, and of a breaking of the scale symmetry at the scale 
transition $\tau$. As we shall see, in what follows, $\tau$ will be 
identified with the de Broglie scale of the system ($\tau=\hbar/E$), 
since $V \approx v$, when $dt \gg \tau$ (classical behavior), and 
$V \approx w$, when $dt \ll \tau$ (fractal behavior). Recalling that 
$D = 2$ plays the role of a critical dimension \cite{LN93}, we stress that, 
in the asymptotic scaling domain, 
$w \propto (dt/\tau)^{-1/2}$, in agreement with Ref. \cite{FH65} for 
quantum paths, which allows to identify the fractal domain with the 
quantum one. In the present paper, only these simplest scale laws with a 
fractal dimension $D=2$ are considered. As recalled in 
Sec.~\ref{ss:galsr}, 
these laws can be identified with ``Galilean'' approximations of more 
general scale-relativistic laws where the fractal dimension becomes 
itself variable with scale. Therefore, all the results that will be obtained 
here (Schr\"odinger, Klein-Gordon and Dirac equations) are open to new 
generalizations, which we shall detail in forthcoming works.\\

The above description strictly applies to an individual fractal trajectory. 
Now, one of the geometric consequences of the non-differentiability and of 
the subsequent fractal character of space/space-time itself (not only of the 
trajectories) is that there is an infinity of fractal geodesics relating 
any couple of points of this fractal space. It has therefore been suggested 
\cite{LN89} that the description of a quantum mechanical particle, including 
its property of wave-particle duality, could be reduced to the geometric 
properties of the set of fractal geodesics that corresponds to a given state 
of this ``particle''. In such an interpretation, we do not have to endow 
the ``particle'' with internal properties such as mass, spin or charge, 
since the ``particle'' is not identified with a point mass which would 
follow the geodesics, but its internal properties can simply be 
defined as geometric properties of the fractal geodesics themselves. As a 
consequence, any measurement is interpreted as a sorting out of the 
geodesics of which the properties correspond to the resolution scale of 
the measuring device (as an example, if the ``particle" has been observed 
at a given position with a given resolution, this means that the geodesics 
which pass through this domain have been selected) \cite{LN93,LN89}. \\

\begin{figure}
   \centering
\includegraphics{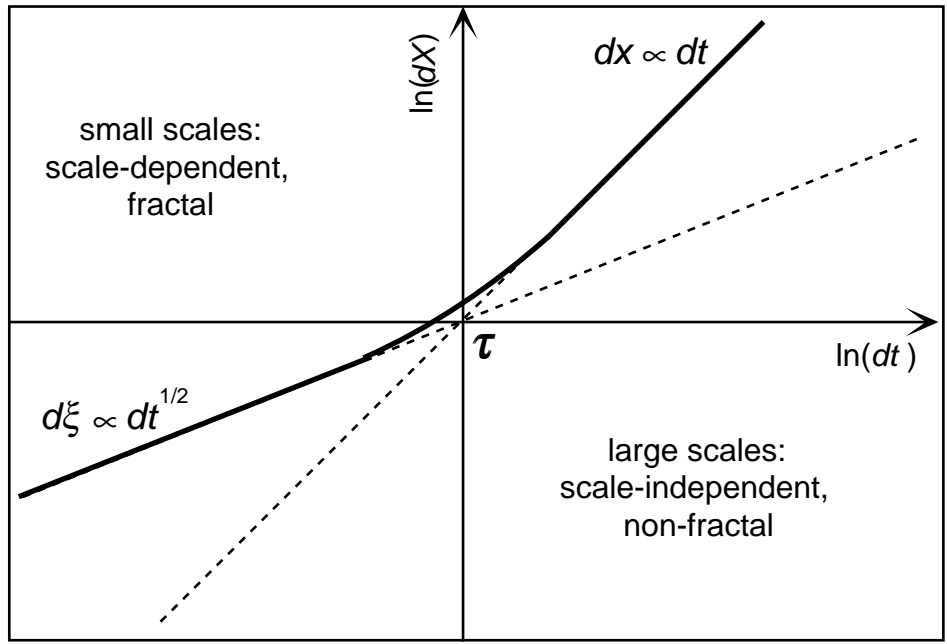}

   \caption{Small-scale to large-scale transition, differentials}
              \label{fig.2}
    \end{figure}

\begin{figure}
   \centering
\includegraphics{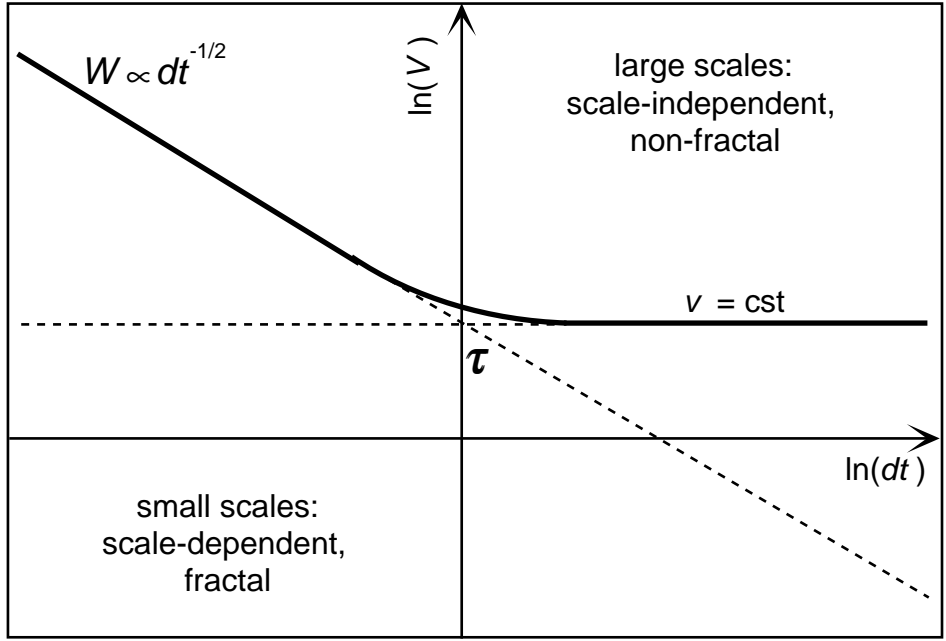}

   \caption{Small-scale to large-scale transition, velocity}
              \label{fig.3}
    \end{figure}

\begin{figure}
   \centering
\includegraphics{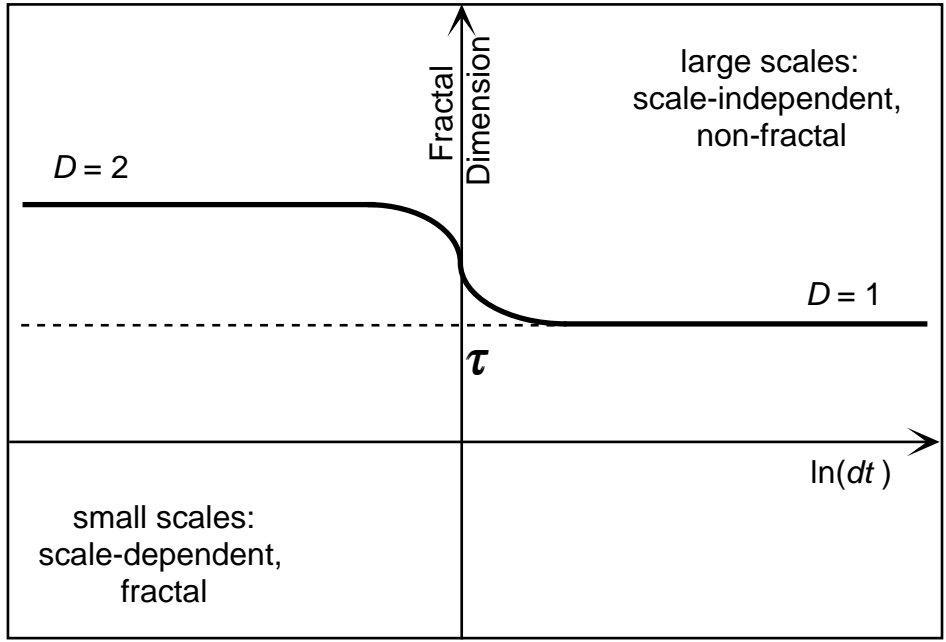}

   \caption{Small-scale to large-scale transition, fractal dimension}
              \label{fig.4}
    \end{figure}

The transition scale appearing in Eq.~(\ref{eq.17}) yields two 
distinct behaviors of the system (particle) depending on the resolution at 
which it is considered (see Figs.~\ref{fig.2}-\ref{fig.4}). 
Equation (\ref{eq.17}) multiplied by $dt$ gives the 
elementary displacement, $dX$, of the system as a sum of two terms
\be
dX = dx + d\xi,
\label{eq.18}
\ee
$d\xi$ representing the fractal fluctuations around the ``classical'' or 
``large-scale'' value, $dx$, each term being defined as
\be
dx = v \; dt,
\label{eq.19}
\ee
\be
d\xi=a \sqrt{2 \cal{D}} (dt^{2})^{1/2D},
\label{eq.20}
\ee
which becomes, for $D=2$,
\be
d\xi=a \sqrt{2 \cal{D}} dt^{1/2},
\label{eq.20bis}
\ee
with $2{\cal D}=\tau _0=\tau v^2$. Owing to Eq.~(\ref{eq.17}), we 
identify $\tau$ as being the Einstein transition scale, $\hbar / E = 
\hbar / {1\over 2} mv^2$, and therefore, as we shall see further on, 
$2{\cal D}=\tau _0$ is a scalar quantity which can be identified with the 
Compton scale, $\hbar / mc$, i.e., it gives the mass of the particle up to 
fundamental constants. \\

We note, from Eqs.~(\ref{eq.18}) to (\ref{eq.20bis}), that $dx$ scales 
as $dt$, while $d\xi$ scales as $dt^{1\over 2}$. Therefore,
the behavior of the system is dominated by the $d\xi$ term in the 
non-differentiable ``small-scale'' domain (below the transition scale), and by 
the $dx$ one in the differentiable ``large-scale'' domain (above the 
transition scale). The logarithmic 
dependence of $dx$ and $d\xi$ with respect to $dt$ is illustrated in 
Fig.~\ref{fig.2}. 

\vspace{5mm}

\vspace{5mm}

\vspace{5mm}

Now, the Schr\"odinger, Klein-Gordon and Dirac equations give results 
applying to measurements performed on quantum objects, but realised with 
classical devices, in the differentiable ``large-scale'' domain. The 
microphysical scale at which the physical systems under study are considered 
induces the sorting out of a bundle of geodesics, corresponding to the scale 
of the systems (see above), while the measurement process implies a smoothing 
out of the geodesic bundle coupled to a 
transition from the non-differentiable ``small-scale'' to the differentiable 
``large-scale'' domain. We are therefore led to define an operator 
$\overline{LS}<\quad>$, 
which we apply to the fractal variables or functions each time we 
are drawn to the ``large-scale'' domain where the $dt$ behavior dominates. The 
effect of $\overline{LS}$ is to extract, from the fractal variables or 
functions to 
which it is applied the ``large-scale part'', i.e., the part scaling as $dt$. 
This justify our above writting $\overline{LS}<\zeta> = 0$ and 
$\overline{LS}<\zeta^2> = 1$, as it is straightforward, 
from Eq.~(\ref{eq.17}), that the terms involving 
$\zeta$ scale as $dt^{1/2}$ and those involving $\zeta^2$ scale as 
$dt$. This also leads us to state, for the $a$ dimensionless coefficient 
in Eq.~(\ref{eq.20bis}), $\overline{LS}<a>=0$ and 
$\overline{LS}<a^2>=1$. \\

Please note the improvement of our new definition in terms of the large-scale 
part with respect to the previous interpretation in terms of an averaging 
process \cite{LN93}. 

\subsection{Differential-time symmetry breaking}
\label{ss:difftsb}

Another consequence of the non-differentiable nature of space (space-time) is 
the breaking of local differential (proper) time reflection invariance. The 
derivative with respect to the time $t$ of a differentiable function $f$ can 
be written twofold
\be
\frac{df}{dt} = \lim_{dt \rightarrow 0}\frac{f(t+dt) - f(t)}{dt} = 
\lim_{dt \rightarrow 0}\frac{f(t) - f(t-dt)}{dt} \; .
\label{eq.21}
\ee

The two definitions are equivalent in the differentiable case. One passes 
from one to the other by the transformation $dt \leftrightarrow -dt$ (local 
differential time reflection invariance), which is therefore an implicit 
discrete symmetry of differentiable physics. In the non-differentiable 
situation, both definitions fail, since the limits are no longer defined. 
In the new framework of scale relativity, the physics is related to the 
behavior of the function during the ``zoom'' operation on the time resolution 
$\delta t$, identified with the differential element $dt$. Two functions 
$f'_+$ and $f'_-$ are therefore defined as explicit functions of $t$ and $dt$
\be
f'_+(t,dt) = \frac{f(t+dt,dt)-f(t,dt)}{dt} \; ,
\label{eq.22}
\ee
\be
f'_-(t,dt) = \frac{f(t,dt)-f(t-dt,dt)}{dt} \; .
\label{eq.23}
\ee

When applied to the space coordinates, these definitions yield, in the 
non-differentiable domain, two velocities that are fractal functions of the 
resolution, $V_+[x(t),t,dt]$ and $V_-[x(t),t,dt]$. In order to go back to the 
``large-scale'' domain and derive the ``large-scale'' velocities appearing in 
Eq.~(\ref{eq.19}), we smooth out each fractal geodesic in the bundles 
selected by the zooming process with balls of radius larger than $\tau$. 
This amounts to carry out a transition from the non-differentiable to the 
differentiable domain and leads to define two ``large-scale'' velocity fields 
now resolution independent: 
$V_+[x(t),t,dt>\tau] = \overline{LS}<V_+[x(t),t,dt]>=v_+[x(t),t]$ and 
$V_-[x(t),t,dt>\tau] = \overline{LS}<V_-[x(t),t,dt]>=v_-[x(t),t]$. The 
important 
new fact appearing here is that, after the transition, there is no more 
reason for these two velocities to be equal. While, in standard mechanics, 
the concept of velocity was one-valued, we must introduce, for the case of a 
non-differentiable space, two velocities instead of one, 
even when going back to the ``large-scale'' domain. This two-valuedness of the 
velocity vector finds its origin in a breaking of the discrete 
time reflection invariance symmetry ($dt \leftrightarrow -dt$). Note that this 
symmetry is actually independent of the time reflection symmetry $T$ 
($t \leftrightarrow -t$) of which the breaking will 
be considered, together with the breaking of the parity $P$ 
($r \leftrightarrow -r$) and of the differential space symmetry 
($dr \leftrightarrow -dr$), in Secs. \ref{s:kgeq} and \ref{s:dieq}. \\

Therefore, if one reverses the sign of the time differential element, 
$v _{+}$ becomes $v _{- }$. But the choice of a reversed time element 
($- dt$) must be as qualified as the initial choice ($dt$) for the 
description of the laws of nature, and we have no way, at this level of the 
description, to define what is the `` past''  and what is the ``future''. The 
only solution to this problem is to consider both the forward $ (dt > 0)$ 
and backward $ (dt < 0) $ processes on the same footing. Then the information 
needed to describe the system is doubled with respect to the standard 
differentiable description. \\

A simple and natural way to account for this doubling consists in using 
complex numbers and the complex product. As we recall hereafter, this 
is the origin of the complex nature of the wave function of 
quantum mechanics, since this wave function can be identified with 
the exponential of the complex action that is naturally introduced in this 
framework. The choice of complex numbers to represent the two-valuedness of 
the velocity is not an arbitrary choice. The use of these numbers is the 
most natural way to generalize to two dimensions the set of real 
numbers. In the same way, as we shall see in Sec. \ref{s:kgeq}, the use of 
bi-quaternionic numbers is the most natural way to generalize them 
to eight dimensions (see Appendix \ref{a:cqn}). \\

\subsection{Covariant derivative operator}
\label{ss:scderiv}

Finally, we describe, in the scaling domain, the elementary displacements for 
both processes, $dX_\pm$, as the sum of a large-scale part, 
$dx_\pm = v_\pm\;dt$, and a fluctuation about this large-scale part, 
$d\xi_\pm$, which is, by definition, of zero large-scale part, 
$\overline{LS}<d\xi_\pm> = 0$ 
\begin{eqnarray} 
  dX_+(t) = v_+\;dt + d\xi_+(t) , \nonumber \\
  dX_-(t) = v_-\;dt + d\xi_-(t) .
\label{eq.24}
\end{eqnarray}

Considering first the large-scale displacements, large-scale forward and 
backward derivatives, $d/dt_+$ and $d/dt_-$, are defined, using the 
large-scale part extraction procedure. Applied to the position 
vector, $x$, they yield the forward and backward large-scale velocities
\be
\frac{d}{dt_+}x(t)=v_+ \; , \qquad \frac{d}{dt_-}x(t) = v_- \; .
\label{eq.25}
\ee

As regards the fluctuations, the generalization to three dimensions of the 
fractal behavior of Eq.~(\ref{eq.20}) writes (for $D=2$)
\be
\overline{LS}<d\xi_{\pm i}\;d\xi_{\pm j}> = \pm 2 \; {\cal D} \; 
\delta _{ij} \; dt \qquad i,j=x,y,z,
\label{eq.26}
\ee
as the $d\xi(t)$'s are of null large-scale part and mutually independent. 
The Kr\"onecker symbol, $\delta _{ij}$, in Eq.~(\ref{eq.26}), implies indeed 
that the large-scale part of every crossed product 
$d\xi_{\pm i}\;d\xi_{\pm j}$, with $i\neq j$, is null. This is due to the fact 
that, even if each term in the product scales as $dt^{1/2}$, each 
behaves as an independent fractal fluctuation around its own large-scale 
part. Therefore, when we proceed to the smoothing out of the geodesic bundle 
during the transition from the small to the large-scale domain, we apply a 
process which is mathematically (not physically) equivalent to a stochastic 
``Wiener'' process, and also more general, since we do not need any Gaussian 
distribution assumption. Thus, we can apply to the large-scale part of the 
$d\xi(t)$ 
product the property of the product of two independent stochastic variables: 
i.e., the large-scale part of the product is the product of the large-scale 
parts, and therefore here zero. \\

To recover local differential time reversibility in terms of a new complex 
process \cite{LN93}, we combine the forward and backward derivatives to 
obtain a complex derivative operator
\be
\frac{\dfr}{dt} = {1\over 2} \left( \frac{d}{dt_+} + \frac{d}{dt_-} \right) 
- {i\over 2} \left(\frac{d}{dt_+} - \frac{d}{dt_-}\right) \; .
\label{eq.27}
\ee

Applying this operator to the position vector yields a complex velocity 
\be
{\cal V} = \frac{\dfr}{dt} x(t) = V -i U = \frac{v_+ + v_-}{2} - i 
\;\frac{v_+ - v_-}{2} \; .
\label{eq.28}
\ee

The minus sign in front of the imaginary term is chosen here in order to 
obtain the Schr\"odinger equation in terms of $\psi$. The reverse choice 
would give the Schr\"odinger equation for the complex conjugate of the wave 
function $\psi^{\dag}$, and would be therefore physically equivalent. \\

The real part, $V$, of the complex velocity, ${\cal V}$, represents the 
standard 
classical velocity, while its imaginary part, $U$, is a new quantity arising 
from non-differentiability. At the usual classical limit, $v_+ = v_- = v$, 
so that $V=v$ and $U = 0$. \\

Contrary to what happens in the differentiable case, the total derivative 
with respect to time of a fractal function $f(x(t),t)$ of integer fractal 
dimension contains finite terms up to higher order \cite{AE05} 
\be
{df\over {dt}}= {\partial f \over {\partial t}} + {\partial f \over 
{\partial x_i}}{dX_i \over {dt}} + {1\over 2}{\partial^2 f \over {\partial 
x_i \partial x_j}}{dX_idX_j\over {dt}} + {1\over 6}{\partial^3 f \over 
{\partial x_i \partial x_j \partial x_k}}{dX_idX_jdX_k\over {dt}} + ...
\label{eq.29}
\ee

Note that it has been shown by Kolwankar and Gangal \cite{KG98} that, if the 
fractal dimension is not an integer, a fractional Taylor expansion can also 
be defined, using the local fractional derivative. \\

In our case, a finite contribution only proceeds from terms of $D$-order, 
while lesser-order terms yield an infinite contribution and higher-order ones 
are negligible. Therefore, in the special case of a fractal dimension $D=2$, 
the total derivative writes
\be
{df\over {dt}} = \frac{\partial f}{\partial t} + \nabla f . {dX\over {dt}} + 
\frac{1}{2} \frac{\partial ^2 f}{\partial x_i \partial x_j} {dX_i dX_j 
\over {dt}} \; .
\label{eq.30}
\ee

Usually the term $dX_i dX_j /dt$ is infinitesimal, but here its large-scale 
part reduces to $\overline{LS}<d \xi_i\; d\xi_j>/dt$. Therefore, thanks to 
Eq.~(\ref{eq.26}), the last term of the large-scale part of Eq.~(\ref{eq.30}) 
amounts to a Laplacian, and we can write  
\be
{df\over {dt_\pm}} = \left({\partial \over {\partial t}} + v_\pm . \nabla 
\pm {\cal D} \Delta\right) f \; .
\label{eq.31}
\ee

Substituting Eqs.~(\ref{eq.31}) into Eq.~(\ref{eq.27}), we finally 
obtain the expression for the complex time derivative operator \cite{LN93}
\be
\frac{\dfr}{dt} = \frac{\partial}{\partial t}Ñ + {\cal V}. \nabla - 
i {\cal D} \Delta \; .
\label{eq.32}
\ee

The passage from standard classical (everywhere differentiable) mechanics 
to the new 
non-differentiable theory can now be implemented by a unique prescription: 
replace the standard time derivative $d/dt$ by the new complex operator 
$\dfr/dt$ \cite{LN93}. In other words, this means that $\dfr/dt$ plays the 
role of a ``covariant derivative operator'' (in analogy with the 
covariant derivative $D_jA^k =  \partial_jA^k + \Gamma_{jl}^k A^l$ replacing 
$\partial_jA^k$ in Einstein's general relativity).

\subsection{Covariant mechanics induced by scale laws}
\label{ss:scmech}

Let us now summarize the main steps by which one may generalize the standard 
classical mechanics using this covariance. We assume that the large-scale 
part of any 
mechanical system can be characterized by a Lagrange function 
${\cal L} (x, {\cal V}, t)$, from which an action ${\cal S}$ is defined
\be
{\cal S} = \int_{t_1}^{t_2} {\cal L} (x, {\cal V}, t) dt.
\label{eq.33}
\ee

Our Lagrange function and action are a priori complex and are obtained from 
the classical Lagrange function $L (x, dx/dt, t)$ and classical action $S$ 
precisely from applying the above prescription $d/dt \rightarrow \dfr/dt$. 
The action principle, which is a stationary action principle applied on 
this new complex action, leads to the generalized Euler-Lagrange equations 
\cite{LN93}
\be
\frac{\dfr}{dt} \frac{\partial {\cal L}}{\partial {\cal V}_i} = 
\frac{\partial {\cal L}}{\partial x_i} \; ,
\label{eq.34}
\ee
which are exactly the equations one would have obtained from applying the 
covariant derivative operator ($d/dt \rightarrow \dfr/dt$) to the usual 
classical Euler-Lagrange equations themselves: this result demonstrates 
the self-consistency of the approach and vindicates the use of complex 
numbers (see Appendix \ref{a:cqn}). Other fundamental results of standard 
classical mechanics are also 
generalized in the same way. In particular, assuming homogeneity of space 
in the mean leads to define a generalized complex momentum and a 
complex energy given by
\be
{\cal P} = \frac{\partial {\cal L}}{\partial {\cal V}},\qquad {\cal E} = 
{\cal P} {\cal V} - {\cal L}.
\label{eq.35}
\ee

If we now consider the action as a functional of the upper limit of 
integration in Eq.~(\ref{eq.33}), the variation of the action from a 
trajectory to another nearby trajectory, when combined with 
Eq.~(\ref{eq.34}), yields a generalization of other well-known relations 
of standard mechanics
\begin{equation}
{\cal P} = \nabla {\cal S} , \qquad {\cal E} = -\partial {\cal S}/\partial t.
\label{eq.36}
\end{equation}

\subsection{Generalized Newton-Schr\"odinger Equation}
\label{ss:gnscheq}

Let us now specialize our study, and consider Newtonian mechanics, i.e., 
the general case when the structuring external field is a scalar potential, 
$\Phi$. The Lagrange function of a closed system, $L=\frac{1}{2}mv^2 -\Phi$, 
is generalized, in the large-scale domain, as ${\cal L} (x,{\cal V},t)=
\frac{1}{2}m{\cal V}^2-\Phi$. 
The Euler-Lagrange equations keep the form of Newton's fundamental equation 
of dynamics
\be
m \frac{\dfr}{dt} {\cal V}= - \nabla \Phi ,
\label{eq.37}
\ee
which is now written in terms of complex variables and complex operators. \\

In the case when there is no external field, the covariance is 
explicit, since Eq.~(\ref{eq.37}) takes the form of the equation of inertial 
motion
\be
\dfr {\cal V} /dt = 0.
\label{eq.37bis}
\ee

This can be compared to what happens in 
general relativity, where the equivalence principle of gravitation and 
inertia leads to a strong covariance principle, expressed by the fact that 
one may always find a coordinate system in which the metric is locally 
Minkowskian. This  means that, in this coordinate system, the covariant 
equation of motion of a free particle is that of inertial motion
\be 
Du_{\mu}=0,
\label{eq.37ter}
\ee
where $u_{\mu}$ is the four-velocity of the particle and $D$, 
the covariant derivative operator of general relativity, defined by its 
application to any four-vector $A^{\nu}$ as
\be
D_{\mu}A^{\nu}=\partial_{\mu}A^{\nu} + \Gamma^{\nu}_{\rho \mu}A^{\rho}.
\ee

Written in any coordinate system, Eq.~(\ref{eq.37ter}) becomes the local 
geodesic equation
\be
{d^2x^{\mu}\over {ds^2}} + \Gamma^{\mu}_{\nu \rho}{dx^{\nu}\over {ds}}
{dx^{\rho}\over {ds}} = 0.
\ee

The covariance induced by scale effects leads to an analogous transformation 
of the equation of motions, which, as we show below, become the 
Schr\"odinger, Klein-Gordon and Dirac equations, and  which we are 
therefore allowed to consider as local geodesic equations. \\

In both cases, with or without external field, the complex momentum 
${\cal P}$ reads
\be
{\cal P} = m {\cal V} ,
\label{eq.38}
\ee
so that, from Eq.~(\ref{eq.36}), the complex velocity ${\cal V}$ appears as 
a gradient, namely the gradient of the complex action
\be
{\cal V} = \nabla {\cal S}/ m.
\label{eq.39}
\ee

We now introduce a complex wave function $\psi$ which is nothing but another 
expression for the complex action ${\cal S}$
\be
\psi = e^{i{\cal S}/{\cal S}_{0}}.
\label{eq.40}
\end{equation}

The factor ${\cal S}_{0}$ has the dimension of an action (i.e., an angular 
momentum) and must be introduced at least for dimensional reasons. We show, 
in what follows, that, when this formalism is applied to microphysics, 
$S_0$ is nothing but the $\hbar$ fundamental constant. It is therefore 
related to the complex velocity appearing in Eq.~(\ref{eq.39}) as follows
\be
{\cal V} = - i \, \frac{{\cal S}_{0}}{m} \, \nabla (\ln \psi).
\label{eq.41}
\ee

We have now at our disposal all the mathematical tools needed to write the 
fundamental equation of dynamics of Eq.~(\ref{eq.37}) in terms of the new 
quantity $\psi$. It takes the form
\be
i {\cal S}_{0} \frac{\dfr}{dt}(\nabla \ln \psi) = \nabla \Phi.
\label{eq.42}
\ee

It is worth noting that the inertial mass $m$ has disappeared from 
this equation. \\

Now one should be aware that ${\dfr}$ and $\nabla$ do not commute. However, 
as we shall see in the followings, there is a particular choice of the 
arbitrary constant ${\cal S}_{0}$ for which ${\dfr}(\nabla \ln \psi)/dt$ is 
nevertheless a gradient. \\

Replacing $\dfr/dt$ by its expression, given by Eq.~(\ref{eq.32}), yields
\be
\nabla   \Phi  =  i {\cal S}_{0} \left(\frac{\partial}{\partial t} + {\cal V}. 
\nabla - i {\cal D} \Delta\right) (\nabla \ln \psi),
\label{eq.43}   
\ee
and replacing once again ${\cal V}$ by its expression in Eq.~(\ref{eq.41}), 
we obtain 
\be
\nabla   \Phi  =   i {\cal S}_{0} \left[ \frac{\partial }{\partial t} \nabla   
\ln\psi   - i \left\{  \frac{{\cal S}_{0}}{m} (\nabla   \ln\psi  . \nabla   )
(\nabla   \ln\psi ) + {\cal D} \Delta (\nabla   \ln\psi )\right\}\right] .
\label{eq.44}
\ee

Consider now the remarkable identity
\be
(\nabla \ln f)^{2} + \Delta \ln f =\frac{\Delta f}{f} \; ,
\label{eq.45}
\ee
which proceeds from the following tensorial derivation
\begin{eqnarray}
\partial_{\mu} \partial^{\mu} \ln f +\partial_{\mu} \ln f \partial^{\mu} 
\ln f &=& \partial_{\mu} \frac{\partial^{\mu} f}{f}+\frac{\partial_{\mu} f}
{f}\frac{\partial^{\mu} f}{f} = \frac{f \partial_{\mu} \partial^{\mu} f - 
\partial_{\mu} f \partial^{\mu} f}{f^{2}}+\frac{ \partial_{\mu} f 
\partial^{\mu} f}{f^{2}} \nonumber \\
&=& \frac{\partial_{\mu} \partial^{\mu} f}{f} \; . 
\label{eq.46}
\end{eqnarray}

When we apply this identity to $\psi$ and take its gradient, we obtain
\be
\nabla\left(\frac{\Delta \psi}{\psi}\right)=\nabla[(\nabla \ln \psi)^{2} + 
\Delta \ln \psi] .
\label{eq.47}
\ee

The second term in the right-hand side of this expression can be transformed, 
using the fact that $\nabla$ and $\Delta$ commute, i.e.,
\be
\nabla \Delta =\Delta \nabla , 
\label{eq.48}
\ee
and so can be the first term, thanks to another remarkable identity
\be
\nabla (\nabla f)^{2}=2 (\nabla f . \nabla) (\nabla f) ,
\label{eq.49}
\end{equation}
that we apply to $f=\ln \psi$. We finally obtain
\be
\nabla\left(\frac{\Delta \psi}{\psi}\right)= 2 (\nabla \ln\psi . \nabla )
(\nabla \ln \psi )  + \Delta (\nabla \ln \psi).
\label{eq.50}
\ee

We recognize, in the right-hand side of this equation, the two terms of 
Eq.~(\ref{eq.44}), which were respectively in factor of ${\cal S}_{0}/{m}$ 
and ${\cal D}$. Therefore, the particular choice 
\be
{\cal S}_{0}=2 m {\cal D}
\label{eq.51}
\ee
allows us to simplify the right-hand side of Eq.~(\ref{eq.44}). 
The simplification is twofold: several complicated terms are compacted into 
a simple one and the final remaining term is a gradient, which means that 
the fundamental equation of dynamics can now be integrated in a universal 
way. The wave function in Eq.~(\ref{eq.40}) is therefore defined as
\be
\psi = e^{i{\cal S}/2m{\cal D}},
\label{eq.52}
\ee
and is solution of the fundamental equation of dynamics, Eq.~({\ref{eq.37}), 
which we write
\be
\frac{\dfr}{dt} {\cal V} = -2 {\cal D} \nabla \left\{i \frac{\partial}
{\partial t} \ln \psi + {\cal D} \frac{\Delta \psi}{\psi}\right\} = 
-\nabla \Phi / m.
\label{eq.53}
\ee

Integrating this equation finally yields
\be
{\cal D}^2 \Delta \psi + i {\cal D} \frac{\partial}{\partial t} \psi - 
\frac{\Phi}{2m}\psi = 0,
\label{eq.54}
\ee
up to an arbitrary phase factor which may be set to zero by a suitable 
choice of the $\psi$ phase. \\

We are now able to enlight the meaning of the choice ${\cal S}_{0}=2 m 
{\cal D}$. At first sight, it is not a necessary condition from the 
viewpoint of physics, but merely a simplifying choice as regards the equation 
form. Indeed, it is only under this particular choice that the 
fundamental equation of dynamics can be integrated. However, if we do not 
make this choice, the $\psi$ function is a solution of a third order, non 
linear, complicated equation such that no precise physical meaning can be 
given to it. We therefore claim that our choice ${\cal S}_{0}=2 m {\cal D}$ 
has a profound physical significance, since, in our construction, the 
meaning of $\psi$ is directly related to the fact that it is a solution of the 
below obtained Schr\"odinger equation. \\

In the case of standard quantum mechanics, as applied to microphysics, this 
result shows that there is a natural link between the Compton relation and 
the Schr\"odinger equation. In this case, indeed, ${\cal S}_{0}$ is nothing 
but the fundamental action constant $\hbar$, while ${\cal D}$ defines the 
fractal/non-fractal transition (i.e., the transition from explicit scale 
dependence to scale independence in the rest frame), 
$\lambda={2{\cal D}/c}$. Therefore, the relation ${\cal S}_{0}=2 m {\cal D}$ 
becomes a relation between mass and the fractal to scale independence 
transition, which writes
\be
\lambda_{c}=\frac{\hbar}{mc} \; .
\label{eq.55}
\ee

We recognize here the definition of the Compton lenght. Its profound meaning - 
i.e., up to the fundamental constants $\hbar$ and $c$, that of inertial mass 
itself - is thus given, in our framework by the transition scale from 
explicit scale dependence (at small scales) to scale independence (at large 
scales). We note that this lenght-scale should be understood as an ``object'' 
of scale space, not of standard space. \\

We recover, in this case, the standard form of Schr\"odinger's equation
\be
\frac{\hbar^2}{2m} \Delta \psi + i \hbar \frac{\partial}{\partial t}\psi = 
\Phi \psi .
\label{eq.56}
\ee

The statistical meaning of the wave function (Born postulate) can now be 
deduced from the very construction of the theory. Even in the case of only 
one particle, the virtual geodesic family is infinite (this remains true 
even in the zero particle case, i.e., for the vacuum field). The particle 
properties are assimilated to those of a random subset of the geodesics in 
the family, and its probability to be found at a given position must be 
proportional to the density of the geodesic fluid. This density can easily 
be calculated in our formalism, since the imaginary part of 
Eq.~(\ref{eq.54}) writes in terms of $\rho = \psi \psi^{\dag}$
\be
{\partial\rho\over \partial t} + \mbox{div} (\rho V) = 0,
\label{eq.57}
\ee
where $V$ is the real part of the complex velocity, which is identified, 
at the classical limit, with the classical velocity. This equation is 
recognized as the equation of continuity, implying that 
$\rho = \psi \psi^{\dag}$ represents the fluid density which is proportional 
to the probability density, thus ensuring the validity of Born's postulate. 
The remarkable new feature that allows us to obtain such a result is that 
the continuity equation is not written as an additional a priori equation, 
but is now a part of our generalized equation of dynamics.

\section{Complex Klein-Gordon equation}
\label{s:ckgeq}

In Sec.~\ref{s:schro}, the Schr\"odinger equation has been derived, 
from first principles, as a geodesic equation in a fractal three space, for 
non-relativistic motion in the framework of Galilean scale relativity. In the 
rest of this article we shall be concerned with relativistic motion in the 
same framework of Galilean scale relativity and shall be led to derive the 
corresponding free-particle quantum mechanical equations (Klein-Gordon's 
and Dirac's) as geodesic equations in a four-dimensional fractal 
space-time. \\
 
The Klein-Gordon equation has already been established, in this framework, 
in a complex form \cite{LN96A}. We summarize and update, in the current 
section, the main steps of its derivation and shall be led to give a 
bi-quaternionic version of the same equation in Sec.~\ref{s:kgeq}. 

\subsection{Motion relativistic covariant derivative operator}
\label{ss:mrcdo}

Most elements of the approach summarized in Sec.~\ref{s:schro} remain 
correct in the motion-relativistic case, with the time, $t$, replaced by the 
proper time, $s$, as the curvilinear parameter along the geodesics. Now, not 
only space, but 
the full space-time continuum is considered to be non-differentiable, thus 
fractal. We consider a small increment $dX_{\mu}$ of a non-differentiable 
four-coordinate along one of the geodesics of the fractal space-time. We can, 
as above, decompose $dX^{\mu}$ in terms of a large-scale part 
$\overline{LS}<dX^{\mu}>=dx^{\mu}=v_{\mu}ds$ and a fluctuation respective to 
this large-scale part $d\xi^{\mu}$, such that 
$\overline{LS}<d\xi^{ \mu }>=0$, by definition. \\

As in the non-relativistic motion case, the non-differentiable nature of 
space-time implies the breaking of the reflection invariance at the 
infinitesimal level. If one reverses the sign of the proper time differential 
element, the large-scale part of the velocity  $v _{+}$ becomes $v _{- }$. We 
therefore consider again both the forward 
$ (ds > 0)$ and backward $ (ds < 0) $ processes on the same footing. Then 
the information needed to describe the system is doubled with respect to the 
classical differentiable description. We have seen that this fundamental 
two-valuedness can be accounted for by the use of complex numbers. The new 
complex process, as a whole, recovers the fundamental property of 
microscopic reversibility. \\

One is then led to write the elementary displacement along a geodesic of 
fractal dimension $D=2$, respectively for the forward $(+)$ and backward 
$(- )$ processes, under the form
\begin{equation}
\label{1.}
dX _{\pm } ^{\mu } =  d _{\pm }x ^{\mu }  +  d\xi _{\pm } ^{\mu }  =  
 v _{\pm } ^{\mu } ds  + u _{\pm }^{\mu }\sqrt{2 {\cal D}} ds ^{1/2},   
\end{equation}
with $d _{\pm }x ^{\mu }  = v _{\pm } ^{\mu } d s$ and 
$d\xi_{\pm }^{ \mu}  =u _{\pm }^{\mu }\sqrt{2 {\cal D}} ds ^{1/2}$. In these 
expressions, $u _{\pm } ^{\mu } $ is a dimensionless fluctuation, and the 
length-scale $2 {\cal D}$ is introduced for dimensional purpose. We 
define the large-scale forward and backward derivatives, $d/ds _{+}$ and 
$d/ds_{-}$, using the large-scale part extraction procedure defined in 
Sec.~\ref{ss:scderiv}, as
\be
\label{2.}
\frac{d}{ds _{\pm }} f(s)  =   lim _{\delta s\rightarrow 0\pm} \; 
\overline{LS}\left<\frac{f(s +\delta s ) - f(s)}{\delta s} \right> .   
\ee

Once applied to $x^\mu $, they yield the forward and backward 
large-scale four-velocities 
\begin{equation}
\label{3.}
  \frac{d}{ds _{+}} x ^{\mu  }(s)  = v ^{\mu } _{+}  \qquad 
\frac{d}{ds _{- }} x ^{\mu  }(s)  = v ^{\mu } _{-}. 
\end{equation}

The forward and backward derivatives of Eq.~(\ref{3.}) can be combined to 
construct a complex derivative operator
\begin{equation}
\frac{\dfr}{ds} = {1\over 2} \left( \frac{d}{ds_+} + \frac{d}{ds_-} \right) 
- {i\over 2} \left(\frac{d}{ds_+} - \frac{d}{ds_-}\right) .
\label{4.}
\end{equation}

When applied to the position vector, this operator yields a complex 
four-velocity 
\begin{equation}
\label{5.}
{\cal V}  ^{\mu } =    \frac{\dfr}{ds}  x ^{\mu }   =   V ^{\mu  }- i  
U ^{\mu  } =    \frac{v ^{\mu } _{+ }+ v ^{\mu } _{- }}{2}   -  i    
\frac{v ^{\mu } _{+ }- v ^{\mu } _{- }}{2} \;  .   
\end{equation}

As regards the fluctuations, the generalization to four dimensions of the 
fractal behavior described in Eq.~(\ref{1.}) gives
\begin{equation}
\label{6.}
\overline{LS}<d\xi  ^{\mu } _{\pm }d\xi  ^{\nu } _{\pm }> =  \mp  2{\cal D}  
\eta  ^{\mu \nu } ds   .   
\end{equation}

As noted in Sec.~\ref{ss:scderiv}, each term in the left-hand side 
product behaves as an independent fractal fluctuation around its own 
large-scale part, which justifies a mathematical treatement analogous to that 
of a stochastic Wiener process. We make in the present paper the choice of 
a $(+,-,-,-)$ signature for the Minkoskian metric of the classical space-time 
$\eta  ^{\mu \nu }$. Therefore, the difficulty comes from the fact 
that a diffusion (Wiener) process makes sense only in $R ^{4}$ where the 
``metric'' $\eta  ^{\mu \nu }$ should be positive definite, if one wants to 
interpret the continuity equation satisfied by the probability density as a 
Kolmogorov equation \cite{RH68}. \\

Several proposals have been made to solve this problem. 
Dohrn and Guerra \cite{DG85} introduce the above ``Brownian metric''  and a 
kinetic metric $g _{\mu \nu }$, and obtain a compatibility condition between 
them which reads $g _{\mu \nu } \eta  ^{\mu \varrho } \eta^{\nu \lambda } 
= g ^{\varrho \lambda }$. An equivalent method was developed by Zastawniak 
\cite{TZ90}, who introduces, in addition to the covariant forward and backward 
drifts $v ^{\mu } _{+}$ and $v ^{\mu } _{- }$ (be careful that our notations 
are different from his), new forward and backward drifts $b ^{\mu } _{+}$ 
and $b ^{\mu } _{-}$. In terms of these drifts the Fokker-Planck equations 
(that one can derive from the Klein-Gordon equation) become Kolmogorov 
equations for a standard Markov-Wiener diffusion process in $R ^{4}$. 
Serva \cite{MS88} gives up Markov processes and considers a covariant process 
that belongs to a larger class, known as ``Bernstein processes''. \\ 

All these proposals are equivalent in the end, and amount to transform a 
Laplacian operator in $R ^{4}$ into a Dalembertian. Namely, the two forward 
and backward differentials of a function $f({x},{ s})$ write
\begin{equation}
\label{7.}
{df\over {ds _{\pm }}} =  \left ({\partial \over {\partial s}}  + v ^{\mu } 
_{\pm  } \partial  _{\mu }   \mp   {\cal D}   \partial   ^{\mu }
\partial  _{\mu } \right ) f    .   
\end{equation}

As, in what follows, we only consider $s$-stationary functions, i.e., 
functions that do not explicitly depend on the proper time $s$, 
the complex covariant derivative operator reduces to
\begin{equation}
\label{8.}
\frac {\dfr}{ds}   =    ({\cal V}  ^{\mu }  +  i  {\cal D}  
\partial   ^{\mu } ) \partial  _{\mu }   .        
\end{equation}

The plus sign in front of the Dalembertian comes from the choice of the 
metric signature.

\subsection{Complex stationary action principle}
\label{ss:csap}

Let us now assume that the large-scale part of any mechanical system can be 
characterized by a 
complex action ${\cal S}$. The same definition of the action as in standard 
relativistic motion mechanics allows us to write, according to the stationary 
action principle,
\be
\delta {\cal S}= -mc \; \delta \int^{b}_{a} ds = 0 ,
\label{eq.81}
\ee
between two fixed points $a$ and $b$, provided $ds$ is defined as 
$ds=\overline{LS}<\sqrt{dX^{\nu}dX_{\nu}}>$. Differentiating the 
integrand, we obtain
\be
\delta {\cal S}= -mc \int^{b}_{a} {\cal V}_{\nu} \;d (\delta x^{\nu}).
\label{eq.82}
\ee
with $\delta x^{\nu} = \overline{LS}<\delta X^{\nu}>$. \\

Integrating by parts yields
\be
\delta {\cal S}= - \left [ mc \; \delta x^{\nu} \right ]^{b}_{a} + 
mc \int^{b}_{a} \delta x^{\nu} \;  {d {\cal V}_{\nu} \over {ds}} \; ds.
\label{eq.83}
\ee

To get the equations of motion, one has to determine $\delta {\cal S}=0$ 
between the same two points, i.e., at the limits 
$(\delta x^{\nu})_a = (\delta x^{\nu})_b = 0$. From Eq.~(\ref{eq.83}), 
we therefore obtain a differential geodesic equation
\be
{d {\cal V}_{\nu} \over {ds}}=0.
\label{eq.84}
\ee

We can also write the elementary variation of the action as a functional of 
the coordinates. We have thus to consider the point $a$ as fixed, so that 
$(\delta x^{\nu})_a =0$. The second point $b$ must be considered as variable. 
The only admissible solutions are those which satisfy the 
equations of motion. Therefore, the integral in Eq.~(\ref{eq.83}) vanishes. 
Simply writing $(\delta x^{\nu})_b$ as $\delta x^{\nu}$ gives
\be
\delta {\cal S}= - mc \; {\cal V}_{\nu} \; \delta x^{\nu}.
\label{eq.85}
\ee

The minus sign in the right-hand side of Eq.~(\ref{eq.85}) proceeds 
from our choice of the metric signature (see Sec. \ref{ss:mrcdo}). 
The complex four-momentum can thus be written as
\be
{\cal P}_{\nu}=mc{\cal V}_{\nu}= -\partial_{\nu}{\cal S}.
\label{eq.86}
\ee

Now, the complex action, ${\cal S}$, characterizes completely the dynamical 
state of the particle, and we can introduce a complex wave function
\begin{equation}
\label{12.}
\psi    =   e  ^{i{\cal S} /{\cal S}_0 }.    
\end{equation}

It is linked to the complex four-velocity by Eq.~(\ref{eq.86}), which gives
\be
{\cal V} _{\nu } = {{i {\cal S}_0\over {mc}}} \partial_{\nu }ln\psi.
\label{13.}
\ee

\subsection{Free-particle complex Klein-Gordon equation}
\label{ss:ckg}

We now apply the scale-relativistic prescription: replace the derivative 
in Eq.~(\ref{eq.84}) by its covariant expression 
given by Eq.~(\ref{8.}). Accounting for the expression of the complex 
four-velocity of Eq.~(\ref{13.}) allows to transform Eq.~(\ref{eq.84}) 
into 
\be
-{{\cal S}_0^2\over {m^2c^2}}\partial^{\mu}ln\psi \partial_{\mu}\partial_{\nu}
ln\psi - {{\cal S}_0 {\cal D}\over {mc}}\partial^{\mu}\partial_{\mu}
\partial_{\nu}ln\psi=0.
\label{14.}
\ee

We see that the particular choice, ${\cal S}_0=\hbar=2mc{\cal D}$, analogous 
to the one made and discussed in Sec.~\ref{ss:gnscheq}, allows us, 
once again, to simplify the left-hand side of Eq.~(\ref{14.}), 
provided we make use of the following identity (which generalizes its 
three-dimensional counterpart given in Eq.~(\ref{eq.46}))
\begin{equation}
\label{10.}
\frac{1}{2}  \partial   ^{\nu } _{  }\left(  \frac{\partial _{\mu }\partial 
^{\mu } \psi }{\psi } \right) = \left( \partial _{\mu } ln\psi + 
\frac{1}{2} \partial _{\mu } \right) \partial ^{\mu } \partial^{\nu }ln\psi . 
\end{equation}

Dividing by the constant factor, ${\cal D}^2$, we obtain the equation of 
motion of the free particle under the form
\begin{equation}
\label{11.}
 \partial ^{\nu } \left(  \frac{\partial ^{\mu }\partial 
_{\mu }\psi}{\psi} \right)     =  0   .    
\end{equation}

Therefore, the Klein-Gordon equation (without electromagnetic field)
\begin{equation}
\label{15.}
\partial ^{\mu } \partial _{\mu}\psi + {m ^{2 }c ^{2}\over {\hbar^2}}\psi=0,  
\end{equation}
becomes an integral of motion  of the free particle, provided the integration 
constant is chosen equal to a squared mass term, $m^2c^2/\hbar^2$. \\

The quantum behavior described by this equation and the probabilistic 
interpretation given to $\psi $ is here reduced to the description of a 
free fall in a fractal space-time, in analogy with Einstein's general 
relativity where a particle subjected to the effect of gravitation is 
described as being in free fall in a curved space-time. \\

Moreover, these equations can indeed be considered as ``covariant'', 
since the relativistic quantum equation written in terms of the complex 
derivative $ \dfr / ds$ has the same form as the equation of a relativistic 
macroscopic and free particle written in terms of the usual derivative 
$d / d s$. \\

It is worth noting here that, contrary to the hope expressed in Ref. 
\cite{LN94B}, the metric form of special and general relativity, 
$V ^{\mu } V _{\mu  }= 1 $, is not conserved in quantum mechanics. Indeed, 
it has been shown by Pissondes \cite{JCP99B} that the free particle 
Klein-Gordon equation, when it is expressed in terms of 
the complex four-velocity ${\cal V} ^{\mu }$, leads to the new following 
equality
\begin{equation}
\label{16.}
{\cal V}  ^{\mu } {\cal V}  _{\mu  }  +  2i  {\cal D}  \partial ^{\mu  }
{\cal V}  _{\mu  }    =    1     .  
\end{equation}

In the scale-relativistic framework, this expression defines the metric that 
is induced by the internal scale structures of the fractal space-time. \\

In the absence of an electromagnetic field, ${\cal V}^{\mu }$  and ${\cal S}$ 
are related by Eq.~(\ref{eq.86}), which we can write
\begin{equation}
\label{17.}
{\cal V} _{\mu }  =   -  \frac{1}{mc}   \partial   _{\mu } {\cal S}    ,   
\end{equation}
so that Eq.~(\ref{16.}) becomes
\begin{equation}
\label{18.}
\partial   ^{\mu }{\cal S}   \partial  _{\mu  }S  -2i   mc {\cal D} 
\partial ^{\mu  } \partial  _{\mu}{\cal S}   =     m ^{2}c^{2 } ,      
\end{equation}
which is the new form taken by the Hamilton-Jacobi equation. \\

Another important property, which comes to light in the above derivation but 
was hidden in Ref. \cite{LN96A}, is the way the characteristic length of the 
fractal (quantum) to differentiable (classical) transition (the Compton 
length of the particle in its rest frame) and the mass term in the 
Klein-Gordon equation appear. 
Here, as in the non-relativistic motion case (Schr\"odinger), we identify the 
transition scale $2{\cal D}$ to the Compton length $\hbar/mc$, provided 
${\cal S}_0$ is the fundamental action constant $\hbar$, in order to obtain 
the motion equation under 
the form of a vanishing four-gradient, as in Eq.~(\ref{11.}). Should we 
make another choice, we would not be able to integrate this equation and 
would obtain a third order, non linear, complicated one to which no precise 
physical meaning could be given. As for the mass term in the final 
Klein-Gordon equation, it does not appear to proceed from a choice of the 
transition length, as it was claimed in Ref. \cite{LN96A}, but to correspond 
to a mere integration constant.

\section{Bi-quaternionic Klein-Gordon equation}
\label{s:kgeq}

It has long been known that the Dirac equation naturally proceeds from the 
Klein-Gordon equation when written in a quaternionic form \cite{CL29,AC37}. 
We propose in the current section to introduce naturally a bi-quaternionic 
covariant derivative operator, leading to the definition of a 
bi-quaternionic velocity and wave-function, which we use to derive the 
Klein-Gordon equation in a bi-quaternionic form. This allows us to obtain, 
in Sec.\ref{s:dieq}, the Dirac equation as a mere consequence, since Dirac 
spinors and bi-quaternions are actually two equivalent representations 
of an electron. \\

The quaternionic formalism, as introduced by Hamilton \cite{WH66}, and 
further developed by Conway \cite{AC37,AC45} is recalled in Appendix 
\ref{a:quat}.

\subsection{Symmetry breaking and bi-quaternionic covariant derivative 
operator}
\label{ss:bqdop}

Most of the approach described in Sec.~\ref{s:ckgeq} remains 
applicable. However, the main new features obtained in 
the now studied case proceed from a deeper description of the scale formalism, 
considering the more general case when the peculiar choice of 
an identification of the  differentials and the resolution variables 
(remember the choice explicited in Sec.~\ref{ss:fracsp}, where we have set 
$dt=\delta t$, which corresponds, in the motion relativistic case, to 
$ds=\delta s$) is given up, implying the subsequent breaking of the 
symmetries: \\
$ds\leftrightarrow -ds$ \\
$dx^{\mu}\leftrightarrow -dx^{\mu}$ \\
$x^{\mu}\leftrightarrow -x^{\mu}$ \\

We have already stressed, in Sec. II.C, that differentiability can be 
recovered (at large scales), even when dealing with non-differentiability. 
However, this implies that, for any set of differential equations describing a 
given process, the physical and mathematical descriptions are only coincident 
on a limited scale range. We can say that any consistent mathematical tool 
lives in a description space which is tangent to the physical space and that 
the validity of this tool is therefore limited to a finite scale region 
around the contact point. For the Schr\"odinger and complex Klein-Gordon 
equations the tangent mathematical space is such that $dt=\delta t$ 
(Schr\"odinger) or $ds=\delta s$ (Klein-Gordon). When jumping to 
smaller (higher energy) scales, we are led to give up this peculiar 
choice, and retain, at the Dirac scale, a new mathematical description 
(detailed below) where the differentials and the resolution variables 
do no more coincide. However, we are aware that this description is only 
valid at the scales where the Dirac equation applies without corrections and 
that we shall be led to improve it when going to yet smaller scales. We know 
actually that, at higher energy, the probability amplitudes must become more 
complicated to take into account, e.g., the isotopic spin, the hypercharge, 
etc.; moreover, one must take into account the radiative 
corrections which, in scale relativity, are identified to new scale 
dependent internal structures, through their description in terms of the 
renormalisation group. This scale depending nature of the coincidences 
between the mathematical and physical worlds can be viewed as a truism since 
it is implicit in the ordinary run of physics, but it becomes a key 
character of the scale-relativistic formalism where the choice of the 
differential variables must be explicited at each stage of the theory. \\

In the scaling domain, the four space-time coordinates 
$X^\mu(s, \epsilon_{\mu}, \epsilon_{s})$ are four fractal functions of the 
proper time $s$ and of the resolutions $\epsilon_\mu$ for the coordinate and 
$\epsilon_s$ for the proper time. \footnote{Please note that the 
$\epsilon _{\mu}$'s are not actually vectors but the diagonalized version of 
a tensorial quantity $\epsilon _{\mu \nu}$, formally analogous to an error 
ellipsoid. This point, irrelevant to the present study, will be 
developed in forthcoming works.} We consider the case when, for an 
elementary displacement $dX^\mu$ corresponding to a shift $ds$ in the 
curvilinear parameter, the resolutions verify $\epsilon_{\mu}<dX^{\mu}$ and 
$\epsilon_{s}<ds$, which implies that, at 
a given $X^\mu$, a forward shift $ds$ of $s$ yields a displacement $dX^\mu$ 
of $X^\mu$ and a backward shift $-ds$ produces a displacement $-dX^\mu$, of 
which the amplitudes are not necessarily equal (see Fig.~\ref{fig.1}). \\

 \begin{figure*}
   \centering
\includegraphics{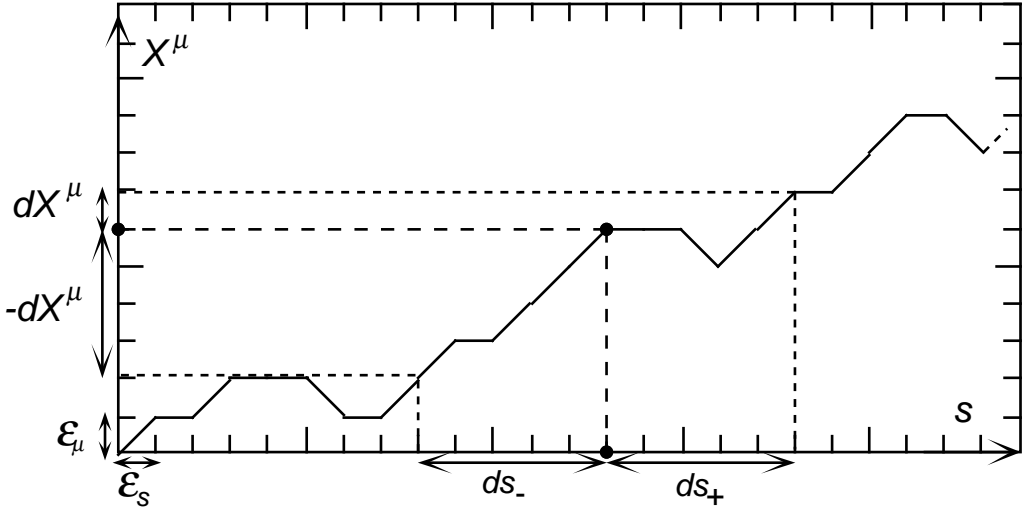}

   \caption{Fractal function $X^{\mu}(s, \epsilon_{\mu}, \epsilon_{s})$}
              \label{fig.1}
    \end{figure*}

We can therefore apply, to these two different elementary displacements, 
the canonical decomposition
\be
dX^{\mu}=dx^{\mu}+d\xi^{\mu},
\label{eq.58}
\ee
with
\be 
\overline{LS}<dX^\mu>=dx^{\mu} = v^{\mu}_{+\atop{\mu}} \; ds,
\label{eq.59}
\ee
\be
d\xi^{\mu}=a^{\mu}_+ \sqrt{2 \cal{D}} (ds^{2})^{\frac{1}{2D}}, \qquad 
\overline{LS}<a^{\mu}_+>=0, \qquad \overline{LS}<(a^{\mu}_+)^2>=1,
\label{eq.60}
\ee
and
\be
-dX^{\mu}=\delta x^{\mu}+\delta \xi^{\mu},
\label{eq.61}
\ee
with
\be 
\overline{LS}<-dX^\mu>=\delta x^{\mu} = v^{\mu}_{-\atop{\mu}} \; ds,
\label{eq.62}
\ee
\be
\delta \xi^{\mu}=a^{\mu}_- \sqrt{2 \cal{D}} (ds^{2})^{\frac{1}{2D}}, \qquad 
\overline{LS}<a^{\mu}_->=0, \qquad \overline{LS}<(a^{\mu}_-)^2>=1.
\label{eq.63}
\ee

In the differentiable case, $dX^\mu=-(-dX^\mu)$, and therefore 
$v^{\mu}_{+\atop{\mu}}=-v^{\mu}_{-\atop{\mu}}$. This is no longer the 
case in the non-differentiable case, where the local symmetry 
$dx^{\mu}\leftrightarrow -dx^{\mu}$ is broken. \\

Furthermore, we must also consider the breaking of the symmetry 
$ds\leftrightarrow -ds$ proceeding from the twofold definition of the 
derivative with respect to the curvilinear parameter $s$. Applied to $X^\mu$, 
considering an elementary displacement $dX^\mu$, the large-scale part 
extraction process gives two large-scale forward and backward 
derivatives $d/ds_+$ and $d/ds_-$, which yield in turn two large-scale 
velocities, which we denote $v^\mu_{{\pm \atop{s}} {+\atop{\mu}}}$. 
Considering now the same extraction process, applied to an elementary 
displacement $-dX^\mu$, the large-scale forward and backward derivatives 
$d/ds_+$ and $d/ds_-$ allow again to define two large-scale velocities, 
denoted $v^\mu_{{\pm \atop{s}} {-\atop{\mu}}}$. We summarize this result as 
\be
v^{\mu}_{{\pm \atop{s}} {+\atop{\mu}}}={dx^\mu\over {ds_{\pm}}}, \qquad 
v^{\mu}_{{\pm \atop{s}} {-\atop{\mu}}}={\delta x^\mu\over {ds_{\pm}}} \; .
\label{eq.64}
\ee

In Galilean scale relativity, with a constant fractal dimension $D=2$, we 
can, at this stage, define several total derivatives with respect 
to $s$ of a fractal function $f$. We write them - using a compact 
straightforward notation with summation over repeated indices - after 
substituting, in the four-dimensional 
analog of Eq.~(\ref{eq.30}), the expressions for the derivatives of the 
$X^\mu$ obtained when using Eqs.~(\ref{eq.58}) to (\ref{eq.63}) with the 
different expressions of Eq.~(\ref{eq.64}) for the large-scale velocities
\be
{df\over {ds}}{}_{{\pm \atop{s}} {\pm \atop{x}} {\pm \atop{y}} 
{\pm \atop{z}} {\pm \atop{t}}}= {\partial f\over {\partial s}} + 
(v^{\mu}_{{\pm \atop{s}} {\pm \atop{\mu}}} + w^{\mu}_{{\pm \atop{s}} 
{\pm \atop{\mu}}}){\partial f\over {\partial X^\mu}} + a^\mu_\pm a^\nu_\pm 
{\cal D}{\partial^2 f\over {dX^\mu dX^\nu}} \; ,
\label{eq.65}
\ee
with
\be
w^\mu=a^\mu {\sqrt {2 {\cal D}}} (ds^2)^{{1\over {2D}}-{1\over 2}}.
\label{eq.66}
\ee

Now, when we apply the large-scale part operator to Eq.~(\ref{eq.65}), using 
Eq.~(\ref{eq.66}) and the 
properties of the $a^\mu_{\pm}$ 's as stated in Eqs.~(\ref{eq.60}) and 
(\ref{eq.63}), the $w^\mu$ 's disappear at the first order, but, at the second 
order, for the fractal dimension $D=2$, we obtain the analogue of 
Eq.~(\ref{6.})
\be
\overline{LS}<w^{\mu}_{{\pm \atop{s}} {\pm \atop{\mu}}} w^{\nu}_
{{\pm \atop{s}} {\pm \atop{\mu}}}>=\mp 2 {\cal D} \eta^{\mu \nu} ds ,
\label{eq.67b}
\ee
the $\mp$ sign in the right-hand side being the inverse of the s-sign in 
the left-hand side. \\

We can therefore write the large-scale part of the total derivatives of 
Eq.~(\ref{eq.65}) as
\be
{df\over {ds}}{}_{{\pm \atop{s}} {\pm \atop{x}} {\pm \atop{y}} 
{\pm \atop{z}} {\pm \atop{t}}}= \left({\partial \over {\partial s}} + 
v^{\mu}_{{\pm \atop{s}} {\pm \atop{\mu}}} {\partial _\mu} 
\mp {\cal D}{\partial^\mu \partial_\mu}\right) f \; ,
\label{eq.67c}
\ee
where the $\mp$ sign in the right-hand side is still the inverse of the 
s-sign. \\

When we apply these derivatives to the position vector $X^\mu$, we obtain, as 
expected,
\be
{dX^{\mu}\over {ds}}{}_{{\pm \atop{s}} {\pm \atop{\mu}}} = v^{\mu}_{{\pm 
\atop{s}} {\pm \atop{\mu}}} .
\label{eq.67}
\ee

We consider now the four fractal functions $-X^\mu(s, \epsilon_\mu, 
\epsilon_s)$. At this description level, there is no reason for 
$(-X^{\mu})(s, \epsilon_\mu, \epsilon_s)$ to be everywhere equal to 
$-(X^{\mu})(s, \epsilon_\mu, \epsilon_s)$, owing to a local 
breaking of the P (for $\mu = x,y,z$) and T (for $\mu = t$) symmetries. 
Furthermore, as we have stressed above for $X^\mu$, 
at a given $-X^\mu$, a forward shift $ds$ of $s$ yields a displacement 
$d(-X^\mu)$ of $-X^\mu$ and a backward shift $-ds$ produces a displacement 
$-d(-X^\mu)$, with no necessarily equal amplitudes. We can therefore apply, 
to these two different elementary displacements, a decomposition similar to 
the one described in Eqs.~(\ref{eq.58}) to (\ref{eq.63}), i.e.,
\be
d(-X^{\mu})={\tilde d} x^{\mu}+{\tilde d}\xi^{\mu},
\label{eq.68}
\ee
with
\be 
\overline{LS}<d(-X^\mu)>={\tilde d}x^{\mu} = {\tilde v}^{\mu}_{+\atop{\mu}} \; ds,
\label{eq.69}
\ee
\be
{\tilde d}\xi^{\mu}={\tilde a}^{\mu}_+ \sqrt{2 \cal{D}} 
(dt^{2})^{\frac{1}{2D}}, \qquad \overline{LS}<{\tilde a}^{\mu}_+>=0, \qquad 
\overline{LS}<({\tilde a}^{\mu}_+)^2>=1,
\label{eq.70}
\ee
and
\be
-d(-X^{\mu})={\tilde \delta} x^{\mu}+{\tilde \delta} \xi^{\mu},
\label{eq.71}
\ee
with
\be 
\overline{LS}<-d(-X^\mu)>={\tilde \delta} x^{\mu} = {\tilde v}^{\mu}_
{-\atop{\mu}} \; ds,
\label{eq.72}
\ee
\be
{\tilde \delta} \xi^{\mu}={\tilde a}^{\mu}_- \sqrt{2 \cal{D}} 
(dt^{2})^{\frac{1}{2D}}, \qquad \overline{LS}<{\tilde a}^{\mu}_->=0, \qquad 
\overline{LS}<({\tilde a}^{\mu}_-)^2>=1,
\label{eq.73}
\ee

Then, we jump to the large-scale parts and are once more confronted to the 
breaking of the $ds\leftrightarrow -ds$ symmetry. Applied to $-X^\mu$, 
considering an elementary displacement $d(-X^\mu)$, the large-scale forward 
and backward derivatives $d/ds_+$ and $d/ds_-$ give again two large-scale 
velocities, which we denote ${\tilde v}^\mu_{{\pm \atop{s}} {+\atop{\mu}}}$. 
After another large-scale part extraction, considering now an elementary 
displacement $-d(-X^\mu)$, the large-scale forward and backward derivatives 
$d/ds_+$ and $d/ds_-$ yield two other large-scale velocities, denoted 
${\tilde v}^\mu_{{\pm \atop{s}} {-\atop{\mu}}}$. We write in short
\be
{\tilde v}^{\mu}_{{\pm \atop{s}} {+\atop{\mu}}}={{\tilde d}x^\mu\over 
ds_{\pm}}, \qquad {\tilde v}^{\mu}_{{\pm \atop{s}} {-\atop{\mu}}}=
{{\tilde \delta} x^\mu\over {ds_{\pm}}} \; .
\label{eq.74}
\ee

We therefore obtain new different total derivatives with respect to $s$ 
of a fractal function $f$, which we write, as above,
\be
{{\tilde d}f\over {ds}}{}_{{\pm \atop{s}} {\pm \atop{x}} {\pm \atop{y}} 
{\pm \atop{z}} {\pm \atop{t}}}= {\partial f\over {\partial s}} + 
({\tilde v}^{\mu}_{{\pm \atop{s}} {\pm \atop{\mu}}} + {\tilde w}^{\mu}_{{\pm 
\atop{s}} {\pm \atop{\mu}}}){\partial f\over {\partial X^\mu}} + 
{\tilde a}^\mu_\pm {\tilde a}^\nu_\pm {\cal D}{\partial^2 f\over 
{dX^\mu dX^\nu}} \; ,
\label{eq.75}
\ee
with
\be
{\tilde w}^\mu={\tilde a}^\mu {\sqrt {2 {\cal D}}} (ds^2)^{{1\over {2D}}-
{1\over 2}}.
\label{eq.76}
\ee

The large-scale part operator applied to Eq.~(\ref{eq.75}), as done in 
Eq.~(\ref{eq.67c}) for Eq.~(\ref{eq.65}), yields the large-scale total 
derivatives
\be
{{\tilde d}f\over {ds}}{}_{{\pm \atop{s}} {\pm \atop{x}} {\pm \atop{y}} 
{\pm \atop{z}} {\pm \atop{t}}}= \left({\partial \over {\partial s}} + 
{\tilde v}^{\mu}_{{\pm \atop{s}} {\pm \atop{\mu}}} {\partial _\mu} 
\mp {\cal D}{\partial^\mu \partial_\mu}\right) f \; .
\label{eq.77b}
\ee

When we apply these derivatives to the position vector $X^\mu$, we obtain, as 
expected again,
\be
{{\tilde d}X^{\mu}\over {ds}}{}_{{\pm \atop{s}} {\pm \atop{\mu}}} = 
{\tilde v}^{\mu}_{{\pm \atop{s}} {\pm \atop{\mu}}}.
\label{eq.77}
\ee

If we consider the simplest peculiar case when the breaking of the symmetry 
$dx^{\mu}\leftrightarrow -dx^{\mu}$ is isotropic as regards the four 
space-time coordinates (i.e., the signs corresponding to the four $\mu$ 
indices are chosen equal), we are left with eight non-degenerate 
components - four $v^{\mu}_{{\pm \atop{s}} {\pm \atop{\mu}}}$ and  
four ${\tilde v}^{\mu}_{{\pm \atop{s}} {\pm \atop{\mu}}}$ - which can be 
used to define a bi-quaternionic four-velocity. We write
\begin{eqnarray}
{\cal V}^\mu=&&{1\over 2}(v^\mu_{++} + {\tilde v}^\mu_{--})-{i\over 2}
(v^\mu_{++} - {\tilde v}^\mu_{--}) +\left[{1\over 2}(v^\mu_{+-} + 
v^\mu_{-+})-{i\over 2}(v^\mu_{+-} - {\tilde v}^\mu_{++})\right] e_1 
\nonumber \\
+&& \left[{1\over 2}(v^\mu_{--} + {\tilde v}^\mu_{+-})-{i\over 2}(v^\mu_{--} - 
{\tilde v}^\mu_{-+})\right] e_2 + \left[{1\over 2}(v^\mu_{-+} + 
{\tilde v}^\mu_{++})-{i\over 2}({\tilde v}^\mu_{-+} + 
{\tilde v}^\mu_{+-})\right] e_3 .
\label{eq.78}
\end{eqnarray}

The freedom in the choice of the actual expression for ${\cal V}^\mu$ will 
be discussed later. It is constrained fy the following requirements: at the 
limit when $\epsilon_\mu \rightarrow dX^\mu$ and $\epsilon_s 
\rightarrow ds$, every $e_i$-term in Eq.~(\ref{eq.78}) goes to zero, 
and, as ${\tilde v}^\mu_{--} = v^\mu_{-+}$ in this limit, one 
recovers the complex velocity of Eq.~(\ref{eq.28}), 
${\cal V}^\mu=[v^\mu_{++}+v^\mu_{-+}-i(v^\mu_{++}-v^\mu_{-+})]/2$; at the 
classical limit, every term in this equation vanishes, save the real term, 
and the velocity becomes classical, i.e., real: ${\cal V}^\mu=v^\mu_{++}$. \\

The bi-quaternionic velocity thus defined corresponds to a bi-quaternionic 
derivative operator $\ddfr /ds$, similarly defined, and yielding, 
when applied to the position vector $X^\mu$, the corresponding velocity. For 
instance, the derivative operator attached to the velocity in 
Eq.~(\ref{eq.78}) writes
\begin{eqnarray}
{\ddfr \over {ds}}={1\over 2}\left({d\over {ds}}{}_{++}+{{\tilde d}\over 
{ds}}{}_{--}\right) -{i\over 2}\left({d\over {ds}}{}_{++} - {{\tilde d}\over 
{ds}}{}_{--}\right) + \left[{1\over 2}\left({d\over {ds}}{}_{+-} + {d\over 
{ds}}{}_{-+}\right) - {i\over 2}\left({d\over {ds}}{}_{+-} - {{\tilde d}\over 
{ds}}{}_{++}\right)\right] e_1 \nonumber \\
+ \left[{1\over 2}\left({d\over {ds}}{}_{--} + {{\tilde d}\over 
{ds}}{}_{+-}\right) - {i\over 2}\left({d\over {ds}}{}_{--} - {{\tilde d}\over 
{ds}}{}_{-+}\right)\right] e_2 + \left[{1\over 2}\left({d\over {ds}}{}_{-+}+
{{\tilde d}\over {ds}}{}_{++}\right) -{i\over 2}\left({{\tilde d}\over 
{ds}}{}_{-+} + {{\tilde d}\over {ds}}{}_{+-}\right)\right] e_3 . \nonumber \\
\label{eq.79}
\end{eqnarray}

Substituting Eqs.~(\ref{eq.67c}) and (\ref{eq.77b}) into 
Eq.~(\ref{eq.79}), we obtain the expression for the bi-quaternionic 
proper-time derivative operator 
\be
{\ddfr \over {ds}}= [1+e_1+e_2+(1-i)e_3]{\partial \over {\partial s}} + 
{\cal V}^\mu \partial_\mu + i{\cal D} \partial ^\mu \partial _\mu ,
\label{eq.80}
\ee
the $+$ sign in front of the Dalambertian proceeding from the choice of the 
metric signature $(+,-,-,-) $. We keep here, for generalty, the $\partial 
/ \partial s$ term, stressing that it is actually of no use, since the 
various physical functions are not explicitly depending on $s$. 
It is easy to check that this operator, applied to the position vector 
$X^\mu$, gives back the bi-quaternionic velocity ${\cal V}^\mu$ of 
Eq.~(\ref{eq.78}). \\

It is worth noting that the expression we have written for ${\cal V}^\mu$ 
in Eq.~(\ref{eq.78}) is one among the various choices we could have 
retained to define the bi-quaternionic velocity. The main constraint limiting 
this choice is the recovery of the complex and real velocities at the 
non-relativistic motion and classical limits. We also choose ${\cal V}^{\mu}$ 
such as to obtain the third term in the right-hand side of Eq.~(\ref{eq.80}) 
under the form of a purely imaginary Dalembertian, which allows to recover 
an integrable equation of motion. To any bi-quaternionic velocity satisfying 
both prescriptions corresponds a bi-quaternionic derivative operator 
$\ddfr /ds$, similarly defined, and yielding 
this velocity when applied to the position vector $X^\mu$. But, 
whatever the definition retained, the derivative operator keeps 
the same form in terms of the bi-quaternionic velocity ${\cal V}^\mu$, as 
given by Eq.~(\ref{eq.89}). 
Therefore, the different choices allowed for its definition merely correspond 
to different mathematical representations leading to the same physical 
result.

\subsection{Bi-quaternionic stationary action principle}
\label{ss:bqsap}

We now apply the stationary action principle, as developed in 
Sec. \ref{ss:csap}, to an elementary variation of a bi-quaternionic action 
${\cal S}$. The free motion equation issued from this principle has the same 
form as Eq.~(\ref{eq.84}), i.e.,
\be
{d {\cal V}_{\mu} \over {ds}}=0,
\label{eq.84b}
\ee
where ${\cal V}_{\mu}$ is the bi-quaternionic four-velocity, e.g., as defined 
in Eq.~(\ref{eq.78}). \\

The elementary variation of the action, considered as a functional of the 
coordinates, gives once more
\be
\delta {\cal S}= - mc \; {\cal V}_{\mu} \; \delta x^{\mu}.
\label{eq.85b}
\ee

We thus obtain the bi-quaternionic four-momentum, as
\be
{\cal P}_{\mu}=mc{\cal V}_{\mu}= -\partial_{\mu}{\cal S}.
\label{eq.86b}
\ee

We can now introduce a bi-quaternionic wave function, which is 
again a re-expression of the action and which we write
\be
\psi^{-1} \partial_{\mu} \psi = {i\over {c S_0}} \partial_{\mu} {\cal S},
\label{eq.87b}
\ee
using, in the left-hand side, the quaternionic product defined in 
Eq.~(\ref{eq.107}). This gives for the bi-quaternionic four-velocity, as 
derived from Eq.~(\ref{eq.86b}),
\be
{\cal V}_{\mu}=i{S_0\over m} \psi^{-1} \partial_{\mu} \psi.
\label{eq.88b}
\ee 

It is worth stressing here that we could choose, for the definition of the 
wave function in Eq.~(\ref{eq.87b}), a commutated expression in the left-hand 
side, i.e., $(\partial_{\mu} \psi) \psi^{-1}$ instead of $\psi^{-1} 
\partial_{\mu} \psi$. But with this reversed choice, 
owing to the non-commutativity of the quaternionic product, we 
could not obtain the motion equation as a vanishing four-gradient, as in 
Eq.~(\ref{eq.95}). Therefore, we retain the above choice 
as the simplest one, i.e., yielding an equation which can be integrated.

\subsection{Free-particle Klein-Gordon equation}
\label{ss:fpkg}

Our next step will be to apply the scale-relativistic prescription:
replace the standard proper time derivative $d/ds$ by the new bi-quaternionic 
operator $\ddfr /ds$, defined in Eq.~(\ref{eq.80}). As, in what follows, we 
only consider $s$-stationary functions, i.e., functions which do not 
explicitly depend on the proper time $s$, the derivative operator reduces 
to
\be
{\ddfr \over {ds}}={\cal V}^\nu \partial_\nu + i{\cal D} \partial ^\nu 
\partial _\nu.
\label{eq.89}
\ee

Now this expression is independent of the peculiar 
choice retained for the bi-quaternionic form of the four-velocity, as the 
representation dependent term in the right-hand side of Eq.~(\ref{eq.80}) 
has vanished. \\

The equation of motion, Eq.~(\ref{eq.84b}), thus writes 
\be
\left ({\cal V}^\nu \partial_\nu + i{\cal D} \partial ^\nu 
\partial _\nu \right ) {\cal V}_{\mu} = 0.
\label{eq.90}
\ee

We replace ${\cal V}_{\mu}$, respectively ${\cal V}^{\nu}$, by their 
expressions given in Eq.~(\ref{eq.88b}) and obtain 
\be
i{S_0\over m} \left ( i{S_0\over m} \psi^{-1} \partial^{\nu} \psi 
\partial_\nu + i{\cal D} \partial ^\nu \partial _\nu \right ) \left ( 
\psi^{-1} \partial_{\mu} \psi \right ) = 0.
\label{eq.91}
\ee

As in Sec. \ref{ss:gnscheq}, the choice ${\cal S}_0= 2 m {\cal D}$ allows 
us to simplify this equation and get
\be
\psi^{-1} \partial^{\nu} \psi \; \partial_\nu (\psi^{-1}  
\partial_{\mu} \psi) + {1\over 2} \partial ^\nu \partial _\nu (\psi^{-1} 
\partial_{\mu} \psi) = 0.
\label{eq.92}
\ee

The definition of the inverse of a quaternion
\be
\psi \psi^{-1} =  \psi^{-1} \psi = 1,
\label{eq.93}
\ee
implies that $\psi$ and $\psi^{-1}$ commute. But this is not necessarily 
the case for $\psi$ and $\partial _{\mu} \psi^{-1}$ nor for $\psi^{-1}$ and 
$\partial _{\mu} \psi$ and their contravariant counterparts (see Appendix 
\ref{a:quat}). However, when we derive Eq.~(\ref{eq.93}) with respect to 
the coordinates, we obtain
\begin{eqnarray}
\psi \; \partial _{\mu} \psi^{-1} &=& - (\partial _{\mu} \psi) \psi^{-1} 
\nonumber \\
\psi^{-1} \partial _{\mu} \psi &=& - (\partial _{\mu} \psi^{-1}) \psi,
\label{eq.94}
\end{eqnarray}
and identical formulae for the contravariant analogues. \\

Developing Eq.~(\ref{eq.92}), using Eqs.~(\ref{eq.94}) and the property 
$\partial^{\nu}\partial_{\nu} \partial_{\mu} = \partial_{\mu}\partial^{\nu} 
\partial_{\nu}$, we obtain, after some calculations,
\be
\partial_{\mu}[(\partial^{\nu}\partial_{\nu} \psi) \psi^{-1}] = 0.
\label{eq.95}
\ee

We integrate this four-gradient and write
\be
(\partial^{\nu}\partial_{\nu} \psi) \psi^{-1} + C = 0 ,
\label{eq.96}
\ee
of which we take the right product by $\psi$ to obtain 
\be
\partial^{\nu}\partial_{\nu} \psi + C \psi = 0.
\label{eq.97}
\ee

We therefore recognize the Klein-Gordon equation for a free particle with 
a mass $m$ so that $m^2c^2/{\hbar}^2=C$.

\section{Dirac equation}
\label{s:dieq}

We now use a long-known property of the quaternionic formalism, which allows 
to obtain the Dirac equation for a free particle as a mere square root of 
the Klein-Gordon operator (see, e.g., Refs. \cite{CL29,AC37}). We give 
below a pedestrian, but accurate, derivation of this property. \\

We first develop the Klein-Gordon equation, as
\be
{1\over {c^2}}{\partial^2 \psi \over {\partial t^2}} = {\partial^2 \psi \over 
{\partial x^2}} + {\partial^2 \psi \over {\partial y^2}} + 
{\partial^2 \psi \over {\partial z^2}} - {m^2c^2\over {\hbar^2}} \psi.
\label{eq.98}
\ee

Thanks to the property of the quaternionic and complex imaginary units 
$e^2_1=e^2_2=e^2_3=i^2=-1$, we can write Eq.~(\ref{eq.98}) under the form
\be
{1\over {c^2}}{\partial^2 \psi \over {\partial t^2}} = e^2_3{\partial^2 \psi 
\over {\partial x^2}}e^2_2 + ie^2_1{\partial^2 \psi \over {\partial y^2}}i + 
e^2_3{\partial^2 \psi \over {\partial z^2}}e^2_1 + i^2{m^2c^2\over {\hbar^2}} 
e^2_3 \psi e^2_3.
\label{eq.99}
\ee

We now take advantage of the anticommutative property of the quaternionic 
units ($e_ie_j=-e_je_i$ for $i\neq j$) to add to the right-hand side of 
Eq.~(\ref{eq.99}) six vanishing couples of terms which we rearrange to 
obtain 
\begin{eqnarray}
{1\over c}{\partial\over {\partial t}} \left ({1\over c}{\partial \psi 
\over {\partial t}}\right ) &=& e_3 {\partial\over {\partial x}} \left ( 
e_3 {\partial \psi \over {\partial x}}e_2 + e_1 {\partial \psi \over 
{\partial y}}i + e_3 {\partial \psi \over {\partial z}}e_1 - 
i{mc\over \hbar}e_3 \psi e_3 \right )e_2 \nonumber \\
&+& e_1 {\partial\over {\partial y}} \left ( 
e_3 {\partial \psi \over {\partial x}}e_2 + e_1 {\partial \psi \over 
{\partial y}}i + e_3 {\partial \psi \over {\partial z}}e_1 - 
i{mc\over \hbar}e_3 \psi e_3 \right )i \nonumber \\
&+& e_3 {\partial\over {\partial z}} \left ( 
e_3 {\partial \psi \over {\partial x}}e_2 + e_1 {\partial \psi \over 
{\partial y}}i + e_3 {\partial \psi \over {\partial z}}e_1 - 
i{mc\over \hbar}e_3 \psi e_3 \right )e_1 \nonumber \\
&-& i{mc\over \hbar}e_3 \left ( 
e_3 {\partial \psi \over {\partial x}}e_2 + e_1 {\partial \psi \over 
{\partial y}}i + e_3 {\partial \psi \over {\partial z}}e_1 - 
i{mc\over \hbar}e_3 \psi e_3 \right )e_3.
\label{eq.100}
\end{eqnarray}

We see that Eq.~(\ref{eq.100}) is obtained by 
applying twice to the bi-quaternionic wavefunction $\psi$ the operator 
$\partial/c\partial t$ written as
\be
{1\over c}{\partial\over {\partial t}} = e_3 {\partial \over 
{\partial x}}e_2 + e_1 {\partial \over 
{\partial y}}i + e_3 {\partial \over {\partial z}}e_1 - 
i{mc\over \hbar}e_3 ( \quad ) e_3.
\label{eq.101}
\ee

The three first Conway matrices $e_3(\quad)e_2$, $e_1(\quad)i$ and 
$e_3(\quad)e_1$ (see Appendix \ref{a:quat}), figuring in the right-hand side 
of Eq.~(\ref{eq.101}), can be written in the compact form $-\alpha^k$, 
with 
\begin{displaymath}
\alpha^k= \left (
          \begin{array}{cc}
           0 & \sigma_k \\ 
           \sigma_k & 0 
           \end{array} 
           \right ), 
\end{displaymath}
the $\sigma_k$'s being the three Pauli matrices, while the fourth Conway 
matrix
\begin{displaymath}
e_3(\quad)e_3= \left (
          \begin{array}{cccc}
           1 & \quad 0 & \; \; 0 & 0 \\ 
           0 & \quad 1 & \; \; 0 & 0 \\
           0 & \quad 0 & \; \; -1 & 0 \\
           0 & \quad 0 & \; \; 0 & -1 
           \end{array}
           \right ) 
\end{displaymath}
can be recognized as the Dirac $\beta$ matrix. We can therefore write 
Eq.~(\ref{eq.101}) as the non-covariant Dirac equation for a free fermion
\be
{1\over c}{\partial \psi \over {\partial t}} = - \alpha^k{\partial \psi \over 
{\partial x^k}} - i{mc\over \hbar}\beta \psi .
\label{eq.102}
\ee

The covariant form, in the Dirac representation, can be recovered by 
applying $ie_3(\quad)e_3$ to Eq.~(\ref{eq.102}). \\

The isomorphism which can be established between the quaternionic and  
spinorial algebrae through the multiplication rules applying to the Pauli 
spin matrices allows us to identify the wave function $\psi$ 
to a Dirac spinor. Spinors and quaternions are both a representation of 
the SL(2,C) group. See Ref. \cite{PR64} for a detailed discussion of the 
spinorial properties of bi-quaternions.

\section{Conclusion}

Three fundamental motion equations which are merely postulated in standard 
microphysics have been recovered, in the framework of Galilean scale 
relativity, as geodesic equations in a fractal space (Schr\"odinger), then 
in a fractal space-time (free Klein-Gordon and Dirac). 
It is interesting to note how the change from classical to 
quantum non-relativistic motion, then from quantum non-relativistic to 
quantum relativistic motion arises from successive symmetry breakings in 
the fractal geodesic picture. \\

First, the complex nature of the wave function is the 
result of the differential (proper) time symmetry breaking, which is  
the simplest effect arising from the fractal structure of space (space-time). 
It allows a statistical interpretation of this wave function in the form of a 
Born postulate. This interpretation is here no more a postulate, since the 
continuity equation 
we write in above Sec. \ref{ss:gnscheq} is now a part of our derived 
equation of dynamics. At this stage, Galilean scale relativity with a 
complex wave function permits the recovery of both the Schr\"odinger 
(Sec. \ref{s:schro}) and Klein-Gordon (Sec. \ref{s:ckgeq}) equations. \\

To go on with the description of the elementary properties encountered 
in the microphysical world, we have to consider further breakings of 
fundamental symmetries, namely the differential coordinate symmetry 
($dx^{\mu} \leftrightarrow -dx^{\mu}$) breaking and the parity and 
time reversal symmetry breakings. These new breakings provide a four-complex 
component wave function (i.e., a eight component wave function), of which 
the most natural mathematical representation is in term of bi-quaternionic 
numbers (see Appendix \ref{a:cqn}). We therefore obtain the spinorial 
and the particle anti-particle nature of elementary objects which we can 
describe as Dirac spinors. Here, spin arises from the isomorphism between 
the quaternionic and spinorial representations, both of which are different 
representations of the SL(2,C) group. At this stage, the Klein-Gordon 
equation can be recovered from a mere stationary action principle in a 
bi-quaternionic form which naturally yields the free Dirac equation. \\

It is worth stressing that these important results only proceed from 
a very restricted use of the scale-relativistic potentialities. We have, in 
the present work, limited our investigations to the induced effects of 
scale laws on the equations of motion, 
in the framework of dilation laws exhibiting a Galilean group structure, i.e., 
a fractal space-time with a constant fractal dimension $D=2$. As it has been 
explained in the introduction, this is only one of the simplest levels at 
which the scale-relativistic program can be realised. We have also made a 
series of simplifying choices, which we have explained and justified all along 
the derivation procedure. Some of them have been dictated by physical or 
experimental considerations, but others only correspond to special cases, 
provisionally retained, and other possibilities will have to be explored in 
the future. We keep therefore confidence in the fact that the further 
developments of the theory should be able to bring new insights into the 
understanding of the laws of nature. \\

\appendix
\section{Complex numbers and bi-quaternions as a covariant representation}
\label{a:cqn}

\subsection{Introduction}

	If one attempts to trace back the origin of most of the quantum behavior, one finds it in the complex nature of the probability amplitude. Feynman particularly stressed this point \cite{RF65}, noticing that, once the first axiom of quantum mechanics is admitted (i.e., the probability amplitude $\psi$ is complex, it is calculated as $\psi=\psi_{1}+\psi_{2}$ for two alternative channels, and the probability density is given by $\psi \psi^{\dagger}$), quantum laws as well as classical laws can be naturally recovered. Is it possible to ``understand" this axiom? In other words, is it possible to derive it from more fundamental principles, for which an intuitive comprehension would be possible? \\

	It is clear that such an understanding is impossible in the framework of quantum mechanics itself (since these statements are its basic axioms), but must be looked for in an enlarged paradigm. The theory of scale relativity provides us with an extension, both of the foundation of physical theories and of the principle of relativity. So it could be worth asking such questions in its frame of thought. In scale relativity, we extend the founding stones of physics by giving up the hypothesis of space-time differentiability; then we extend the principle of relativity by applying it, not only to motion laws, but also to scale laws. \\ 

	We have recalled in Sec. \ref{ss:difftsb} our demonstration \cite{LN93} that one of the main and simplest consequences of the non-differentiability of space-time is that the velocity vector becomes two-valued. This two-valuedness of the velocity implies a two-valuedness of the Lagrange function, and therefore a two-valuedness of the action $S$. Finally, the wave function is defined as a re-expression for the action, so that it will also be two-valued in the simplest case. \\

But we have, up to now, admitted without justification that this two-valuedness is to be described in terms of complex numbers. Why? This is an important point for the understanding of quantum mechanics, but also from the viewpoint of the meaning of the scale-relativistic approach. Indeed, our equations are not simply a ``pasting" of real and imaginary equations, but involve the complex product from the very beginning of the calculations. In particular, the geodesic equation for a free particle takes the form of the equation of inertial motion
\begin{equation}
\label{appA}
\frac{\dfr^2}{dt^2} x^{k}=0,
\end{equation}
in terms of the ``covariant" complex time derivative  operator
\begin{equation}
\frac{\dfr}{dt} =\frac{\partial}{\partial t} + {\cal V} . \nabla - i {\cal D} \Delta.
\end{equation}

Equation (\ref{appA}) amounts to Schr\"odinger's equation (see main text). Since it corresponds to a second derivative, the complex product has mixed the real and imaginary quantities in a very specific way, so that it is not at all a trivial statement that Schr\"odinger's equation is recovered in the end. \\

	The aim of this appendix is precisely to address this specific question: why complex numbers and why bi-quaternions? As we shall see, the answer is that complex numbers achieve a particular representation of quantum mechanics in terms of which the fundamental equations take their simplest form. Other choices for the representation of the two-valuedness and for the new product are possible, but these choices would give to the Schr\"odinger equation a more complicated form, involving additional terms (although its physical meaning would be unchanged). Bi-quaternions follow as a further splitting, at another level, of the complex numbers.

\subsection{Origin of complex numbers and quaternions in quantum mechanics}
\label{ss:cnq}

	Let us return to the step of our demonstration where complex numbers are introduced. All we know is that each component of the velocity now takes two values instead of one. This means that each component of the velocity becomes a vector in a two-dimensional space, or, in other words, that the velocity becomes a two-index tensor. So let us introduce generalized velocities
\begin{equation}
\label{}
V^{k}_{\sigma}=(V^{k},U^{k})  \;\; ; \;\;  k= 1,2,3   \;\; ; \;\;  \sigma  =-,+ .
\end{equation}

This can be generalized to all other physical quantities affected by the two-valuedness: namely, scalars $A$ of the position space become vectors $A_{\sigma}$ of the new 2D-space, etc... . While the generalization of the sum of these quantities is straighforward, $C^{k}_{\sigma} = A^{k}_{\sigma}+B^{k}_{\sigma}$, the generalisation of the product is an open question, which should be derived from first principles. \\

The problem can be put in a general way: it amounts to find a generalization of the standard product that keeps its fundamental physical properties, or at least that keeps the most important ones (e.g., internal composition law) when some of them cannot escape to be lost (e.g., commutativity when jumping to quaternions). \\

From the mathematical point of view, we are here exactly confronted to the well-known problem of the doubling of algebra (see, e.g., Ref. \cite{MP82}). Indeed, the effect of the symmetry breaking $dt \leftrightarrow -dt$ (or $ds \leftrightarrow -ds$) is to replace the algebra ${\cal A}$ in which the classical physical quantities are defined, by a direct sum of two exemplaries of ${\cal A}$, i.e., the space of the pairs $(a,b)$ where $a$ and $b$ belong to ${\cal A}$. The new vectorial space ${\cal A}^2$ must be supplied with a product in order to become itself an algebra (of doubled dimension). The same problem is asked again when one takes also into account the symmetry breakings $dx \leftrightarrow -dx$ and $x^{\mu} \leftrightarrow -x^{\mu}$, which is the aim of the present paper: this leads to new algebra doublings. The mathematical solution to this problem is well-known: the standard algebra doubling amounts to supply ${\cal A}^2$ with the complex product. Then the doubling ${\rm I\! R^2}$ of ${\rm I\! R}$ is the algebra ${\rm I\! \!\!C}$ of complex numbers, the doubling ${\rm I\! \!\!C^2}$ of ${\rm I\!\! \!C}$ is the algebra ${\rm I\! H}$ of quaternions, the doubling ${\rm I\! H^2}$ of quaternions is the algebra of Graves-Cayley octonions. The problem with algebra doubling is that the iterative doubling leads to a progressive deterioration of the algebraic properties. Namely, the quaternion algebra is non-commutative, while the octonion algebra is also non-associative. But an important positive result for physical applications is that the doubling of a metric algebra is a metric algebra \cite{MP82}. \\

These mathematical theorems fully justify the use of complex numbers, then of quaternions, in order to describe the successive doublings due to discrete symmetry breakings at the infinitesimal level, which are themselves more and more profound consequences of space-time non-differentiability. \\

However, we give in what follows complementary arguments of a physical nature, which show that the use of the complex product in the first algebra doubling have a simplifying and covariant effect (we use here the word ``covariant" in the original meaning given to it by Einstein \cite{AE16}, namely, the requirement of the form invariance of fundamental equations). \\

In order to simplify the argument, let us consider the generalization of scalar quantities, for which the product law is the standard product in ${\rm I \!R}$. \\

The first constraint is that the new product must remain an internal composition law. We also make the simplifying assumption that it remains linear in terms of each of the components of the two quantities to be multiplied. 
Therefore we consider a general form for a bilinear internal product
\begin{equation}
\label{}
C^{\gamma}= A^{\alpha} \; \Omega_{\alpha \beta}^{\gamma}\; B^{\beta}
\end{equation}
where the matrix $\Omega_{\alpha \beta}^{\gamma}$ is a tensor (similar to the structure constants of a Lie group) that defines completely the new product. \\

The second physical constraint is that we recover the classical variables and the classical product at the classical limit. The mathematical equivalent of this constraint is the requirement that ${\cal A}$ still be a sub-algebra of ${\cal A}^2$. Therefore we identify $a_0 \in {\cal A}$ with $(a_0,0)$ and we set $(0,1)=\alpha$. This allows us to write the new two-dimensional vectors in the simplified form $a=a_{0}+a_{1} \alpha$, so that the product now writes
\begin{equation}
\label{}
c=( a_{0}+a_{1} \alpha)\,( b_{0}+b_{1} \alpha)= a_{0}b_{0}+a_{1}b_{1} \alpha^{2}+(a_{0}b_{1}+a_{1}b_{0})\alpha .
\end{equation}

The problem is now reduced to find $\alpha^2$, i.e., only two $\Omega$ 
coefficients instead of eight
\begin{equation}
\label{}
\alpha^{2}=\omega_{0}+\omega_{1} \alpha.
\end{equation}

Let us now come back to the beginning of our construction. We have introduced two elementary displacements, a forward and a backward one, each of them made of two terms, a large-scale part and a fluctuation (see Eq.~(\ref{eq.24}))
\begin{eqnarray} 
  dX_+(t) = v_+\;dt + d\xi_+(t) , \nonumber \\
  dX_-(t) = v_-\;dt + d\xi_-(t) .
\end{eqnarray}

Therefore, one can define velocity fluctuations $w_{+}=d\xi_{+}/dt$ and $w_{-}=d\xi_{-}/dt$, then a complete velocity in the doubled algebra \cite{LN99}
\begin{equation}
\label{}
{\cal V} + {\cal W} = \left(\frac{v_{+}+v_{-}}{2} - \alpha \, \frac{v_{+}-v_{-}}{2}\right) + \left(\frac{w_{+}+w_{-}}{2}-\alpha\, \frac{w_{+}-w_{-}}{2}\right) \; .
\end{equation}

Note that in terms of standard methods, this writing would be forbidden since the velocity ${\cal W}$ is infinite from the viewpoint of usual differential calculus (it is $\propto dt^{-1/2}$). But we recall that we give meaning to this concept by considering it as an explicit function of the differential element $dt$, which becomes itself a variable. \\

Now, from the covariance principle, the Lagrange function in the Newtonian case should strictly be written:
\begin{equation}
\label{}
{\cal L}= \frac{1}{2} m \; \overline{LS}<({\cal V} + {\cal W})^{2}>=\frac{1}{2} m \; \left(\overline{LS}<{\cal V}^{2}> + \overline{LS}<{\cal W}^{2}>\right)
\end{equation}

We have $\overline{LS}<{\cal W}>=0$, by definition, and $\overline{LS}<{\cal V}{\cal W}>=0$, because they are mutually independent. But what about $\overline{LS}<{\cal W}^{2}>$ ? The presence of this term would greatly complicate all the subsequent developments toward the Schr\"odinger equation, since it would imply a fundamental divergence of non-relativistic quantum mechanics (note however that such a divergence finally happens in relativistic quantum field theories and leads to the renormalisation and renormalisation group approachs). Let us develop it
\begin{eqnarray}
\label{}
4\overline{LS}<{\cal W}^{2}>&=&\overline{LS}<  [(w_{+}+w_{-})-\alpha\, (w_{+}-w_{-})]^2 > \nonumber \\
&=&\overline{LS}<(w_{+}^2+w_{-}^2)(1+\alpha^2)-2\alpha(w_{+}^2-w_{-}^2)+ 2 w_{+}w_{-}(1-\alpha^2)>.
\end{eqnarray}

Since $\overline{LS}<w_{+}^2>=\overline{LS}<w_{-}^2>$ and $\overline{LS}<w_{+}w_{-}>=0$ (they are mutually independent), we finally find that $\overline{LS}<{\cal W}^{2}>$ can only vanish provided
\begin{equation}
\label{}
\alpha^2=-1,
\end{equation}
namely, $\alpha=i$, the imaginary. Therefore we have shown that the choice of the complex product in the algebra doubling plays an essential physical role, since it allows to suppress what would be additional infinite terms in the final equations of motion.

\subsection{Origin of bi-quaternions}

A last point that must be justified is the use of complex quaternions
(bi-quaternions) for describing the new two-valuedness that leads to
bi-spinors and the Dirac equation. One could think that the argument
given in Sec. \ref{ss:cnq} (algebra doubling) leads to use 
Graves-Cayley octonions (and therefore to give up associativity) in the
case of three successive doublings as considered in this paper. However,
these three doublings are not on the same footing from a physical point
of view: \\
(i) The first two-valuedness comes from a discrete symmetry breaking at
the level of the differential invariant, namely, $dt$ in the case of a
fractal space (yielding the Schr\"odinger equation) and $ds$ in the
case of a fractal space-time (yielding the Klein-Gordon equation). This
means that it has an effect on the total derivatives $d/dt$ and $d/ds$.
This two-valuedness is achieved by the introduction of complex 
variables. \\
(ii) The second two-valuedness (differential parity and time reversal 
violation, ``dX''),
which is specifically introduced and studied in the present paper, comes
from a new discrete symmetry breaking (expected from the giving up of
the differentiability hypothesis) on the 
space-time differential element $dx^{\mu} \leftrightarrow -dx^{\mu}$.
It is subsequent to the first two-valuedness, since it has an effect on
the partial derivative $\partial_{\mu}=\partial / \partial x^{\mu}$ that
intervenes in the complex covariant derivative operator, namely,
\begin{equation}
\frac{\dfr}{ds}= ({\cal V}^{\mu}+i {\cal D} \partial^{\mu})\partial_{\mu}.
\end{equation}
(iii) The third two-valuedness is a standard effect of parity (``P") and 
time reversal (``T'') in
the motion-relativistic situation, which is not specific of our approach
and is already used in the standard construction of Dirac spinors. It
does not lead to a real information doubling, since Dirac spinors
still have only four degrees of freedom as Pauli spinors do. \\

Therefore, from the second and third doublings, complex numbers, then
quaternions, can be introduced, which will affect variables which are
already complex due to the first, more fundamental, doubling. This leads 
to the bi-quaternionic tool we use in the present work. \\

Let us conclude this section by noting that these symmetry breakings are
effective only at the level of the underlying description of elementary
displacements (namely, in the non-differentiable fractal space-time).
The effect of introducing a two-valuedness of variables in terms of
double symmetrical processes precisely amounts to recover symmetry in
terms of the bi-process, and therefore in terms of the quantum tools
which are built from it. But reversely, this remark opens a possible way
of investigation for future research about the origin of other features 
characteristic of the microphysical world.

\subsection{Complex representation and covariant Euler-Lagrange equations.}

Let us confirm by another argument that our choice of a representation in terms of complex numbers for the two-valuedness of the velocity (namely, in the minimal situation that leads to the Schr\"odinger equation) is a simplifying and covariant choice. We indeed demonstrate hereafter that the standard form of the Euler-Lagrange equations is conserved when we combine the forward and backward velocities in terms of a unique complex velocity. \\

In a general way, the Lagrange function is expected to be a function of the variables $x$ and their time derivatives $\dot{x}$. We have found that the number of velocity components $\dot{x}$ is doubled, so that we are led to write
\begin{equation}
\label{app1}
L=L(x,\dot{x}_{+}, \dot{x}_{-},t).
\end{equation}

Instead, we have made the choice to write the Lagrange function as
\begin{equation}
\label{app2}
L=L(x,{\cal V},t).
\end{equation}

We now justify this choice by the covariance principle.
Re-expressed in terms of $\dot{x}_{+}$ and $\dot{x}_{-}$, the Lagrange 
function writes
\begin{equation}
L=L\left(x,\frac{1-i}{2} \; \dot{x}_{+} + \frac{1+i}{2} \; \dot{x}_{-},t\right).
\end{equation}

Therefore we obtain
\begin{equation}
\frac{\partial L}{\partial \dot{x}_{+}}=\frac{1-i}{2} \; \frac{\partial L}{\partial {\cal V}} \;\;\; ; \;\;\; \frac{\partial L}{\partial \dot{x}_{-}}=\frac{1+i}{2} \; \frac{\partial L}{\partial {\cal V}},
\end{equation}
while the new covariant time derivative operator writes
\begin{equation}
\frac{\dfr}{dt}=\frac{1-i}{2} \; \frac{d}{dt_{+}}+\frac{1+i}{2} \; \frac{d}{dt_{-}} \; .
\end{equation}

Let us write the stationary action principle in terms of the Lagrange function, as written in Eq. (\ref{app1})
\begin{equation}
\delta S= \delta \int_{t_1}^{t_2} L(x,\dot{x}_{+}, \dot{x}_{-},t) \; dt = 0.
\end{equation}

It becomes
\begin{equation}
 \int_{t_1}^{t_2}\, \left(  \frac{\partial L}{\partial x} \; \delta x +\frac{\partial L}{\partial \dot{x}_{+}} \; \delta  \dot{x}_{+} + \frac{\partial L}{\partial \dot{x}_{-}} \; \delta  \dot{x}_{-}  \right) \, dt = 0.
\end{equation}

Since $\delta  \dot{x}_{+} =d(\delta x)/dt_{+}$ and $\delta  \dot{x}_{-} =d(\delta x)/dt_{-}$, it takes the form
\begin{equation}
 \int_{t_1}^{t_2}\, \left(  \frac{\partial L}{\partial x} \; \delta x +\frac{\partial L}{\partial {\cal V} }\; \left[ \, \frac{1-i}{2} \; \frac{d}{dt_{+}} + \frac{1+i}{2} \; \frac{d}{dt_{-}}\,\right] \,  \delta  x  \right) \, dt = 0,
\end{equation}
i.e.,
\begin{equation}
\int_{t_1}^{t_2}\, \left(      \frac{\partial L}{\partial x} \; \delta x      +     \frac{\partial L}{\partial {\cal V}} \; \frac{\dfr}{dt} \, \delta  x  \right)\, dt = 0.
\end{equation}

Therefore, after the integration by part, the integral reduces to
\begin{equation}
\int_{t_1}^{t_2}\, \left(      \frac{\partial L}{\partial x}    -   \frac{\dfr}{dt} \, \frac{\partial L}{\partial {\cal V}}   \right) \, \delta  x \,  dt = 0.
\end{equation}

Finally the Euler-Lagrange equations write
\begin{equation}
\frac{\dfr}{dt} \, \frac{\partial L}{\partial {\cal V}} = \frac{\partial L}{\partial x}   .
\end{equation}

They take exactly the form one obtains by writing a stationary action principle based on Eq. (\ref{app2}). Moreover, once done the transformation $d/dt \rightarrow \dfr/dt$, this form is nothing but the standard classical form. This result reinforces the identification of our tool with a ``quantum-covariant'' representation, since, as we have shown in previous works and as we recall in Sec. \ref{s:schro}, this Euler-Lagrange equation can be integrated in the form of a Schr\"odinger equation. \\

\section{Quaternionic calculus} 
\label{a:quat}

In this appendix, we summarize our notations for the quaternionic algebra 
and introduce the elementary properties of quaternionic arithmetic and 
analysis which are of use in the present article. \\

\subsection{Definitions and algebraic properties}
\label{ss:defap}

A bi-quaternion $\phi=(\phi_0,\phi_1,\phi_2,\phi_3)$ is an ordered quadruple 
of complex numbers. The $\phi_i$'s are the components of $\phi$. 
The equality of two quaternions is equivalent to the equality of their 
corresponding components: $\phi=\psi$ if and only if $\phi_i=\psi_i$, with 
$i=0,1,2,3$. \\

The multiplication of a bi-quaternion by a complex number $\alpha$, and the 
addition and substraction of quaternions are defined by
\begin{eqnarray} 
\alpha \phi&=&(\alpha \phi_0, \alpha \phi_1, \alpha \phi_2, \alpha \phi_3), 
\nonumber \\
\phi+\psi&=& (\phi_0+\psi_0, \phi_1+\psi_1, \phi_2+\psi_2, \phi_3+\psi_3), 
\nonumber \\
\phi-\psi&=&\phi + (-1)\psi.
\label{eq.103}
\end{eqnarray}

Addition of quaternions is commutative and associative. The null quaternion 
$\phi=0$, i.e., the neutral element for addition, writes $\phi=(0,0,0,0)$. 
Multiplication of quaternions by complex numbers is commutative, associative 
and distributive. The quaternionic product $\phi\psi$ of two quaternions is 
itself a quaternion. The product is distributive and satisfy, for any 
complex number $\alpha$,
\be
(\alpha \phi)\psi=\phi(\alpha \psi)=\alpha(\phi\psi).
\label{eq.104}
\ee

Any quaternion can be decomposed as
\be
\phi= \phi_0 + e_1\phi_1 + e_2\phi_2 + e_3\phi_3,
\label{eq.105}
\ee
where Hamilton's imaginary units $e_i$ satisfy the associative but 
non-commutative algebra 
\be
e_ie_j= -\delta_{ij} + \sum_{k=1}^3\epsilon_{ijk}e_k, \qquad i,j=1,2,3
\label{eq.106}
\ee
where $\epsilon_{ijk}$ is the usual completely antisymmetric three-index 
tensor with $\epsilon_{123}=1$. From these rules, it is easy to establish 
that the product of two arbitrary quaternions is
\begin{eqnarray}
\phi\psi=(\phi_0\psi_0&-&\phi_1\psi_1-\phi_2\psi_2-\phi_3\psi_3, 
          \phi_0\psi_1+\phi_1\psi_0+\phi_2\psi_3-\phi_3\psi_2, \nonumber \\
          \phi_0\psi_2&+&\phi_2\psi_0+\phi_3\psi_1-\phi_1\psi_3,
          \phi_0\psi_3+\phi_3\psi_0+\phi_1\psi_2-\phi_2\psi_1).
\label{eq.107}
\end{eqnarray}
This product is not in general commutative, but is associative.

The adjoint of a quaternion, also named quaternion conjugate or Hamilton 
conjugate, is defined as 
\be
\phi \rightarrow \bar{\phi}=  (\phi_0, - \phi_1, - \phi_2, - \phi_3).
\label{eq.108}
\ee
 
One also defines the scalar product of quaternions
\be
\phi.\psi={1\over 2} (\phi \bar{\psi} + \psi \bar{\phi}) = {1\over 2} 
(\bar{\phi}\psi + \bar{\psi}\phi) = \phi_0\psi_0+\phi_1\psi_1+
\phi_2\psi_2+\phi_3\psi_3.
\label{eq.109}
\ee

The norm of a quaternion is therefore
\be
\phi. \bar{\phi}=\phi \bar{\phi}=\phi_0^2 + \phi_1^2 +\phi_2^2 +\phi_3^2 
= \bar{\phi}\phi=\phi .\bar{\phi}.
\label{eq.110}
\ee

The norm and scalar product are numbers, in general complex, the norm 
being of course the self-scalar product, but also the self-quaternionic 
product. The norm of a product is the 
product of the norms. When $\phi \bar{\phi}=0$, $\phi$ is said to be null 
or singular. If  $\phi \bar{\phi}=1$, $\phi$ is unimodular. \\

The reciprocal or inverse of any non null quaternion $\phi$ is defined as
\be
\phi^{-1}={\bar{\phi}\over {\phi \bar{\phi}}} \; .
\label{eq.111}
\ee

It possesses the properties:
\begin{eqnarray}
(\phi\psi)^{-1}&=&\psi^{-1}\phi^{-1}, \nonumber \\
\phi\phi^{-1}&=&\phi^{-1}\phi=1.
\label{eq.112}
\end{eqnarray} \\

The latter property is a key tool of our derivation of the bi-quaternionic 
Klein-Gordon equation in Sec. \ref{ss:fpkg}. \\

\subsection{Quaternions and Conway matrices}
\label{ss:qcm}

A connection between quaternions and the Dirac-Eddington matrices has 
first been brought to light by Conway \cite{AC37,AC45}. It gives rise to 
the definition of a set of sixteen Conway matrices, a subsample of which 
is used in Sec. \ref{s:dieq} to derive Dirac's equation 
from Klein-Gordon's. We give below the main steps of their construction. \\

Let us consider any two quaternions $a$ and $b$, and the transformation
\be
\phi \rightarrow \psi=a\phi b,
\label{eq.113}
\ee
which is linear for the elements $(\phi_0, \phi_1, \phi_2, \phi_3)$ and can 
therefore be written in matrix form
\be
\psi=M(a,b)\phi,
\label{eq.114}
\ee
where $M$ is a 4x4 matrix and $\phi$ and $\psi$ are 4x1 column matrices. 
The notation indicates that $M(a,b)$ is determined by $a$ and $b$. \\

If we now consider a couple of linear transformations and their resultant
\be
\psi=a\phi b, \qquad \chi=c\psi d, \qquad \chi=ca\phi bd,
\label{eq.115}
\ee
which can be written in matrix form
\be
\psi=M(a,b)\phi, \qquad \chi=M(c,d)\psi, \qquad \chi=M(c,d)M(a,b)\phi,
\label{eq.116}
\ee
therefore allowing us to write
\be
M(c,d)M(a,b)=M(ca,bd)
\label{eq.117}
\ee

In particular, for $a=c$ and $b=d$, Eq.~(\ref{eq.117}) becomes
\be
M(a,b)^2=M(a^2,b^2).
\label{eq.118}
\ee

If we substitute $a=b=-1$ in Eq.~(\ref{eq.113}), we obtain the identical 
transformation and are thus allowed to write $M(-1,-1)=I$, which is 
the unit matrix. Hence, by Eq.~(\ref{eq.118}), we have the following 
theorem: ``If $a$ and $b$ are any two quaternions satisfying
\be
a^2=b^2=-1,
\label{eq.119}
\ee
then, $M(a,b)^2=I$. In other words, the matrix generated by $a$ and $b$ 
is a square root of unity.'' \\ 

Conway proposes the suggestive notation
\be
M(a,b)=a(\quad)b.
\label{eq.120}
\ee

Now, any of the four quaternions $e_1, e_2, e_3, i$ satisfies 
Eq.~(\ref{eq.119}). We have therefore the following sixteen matrices, 
each a square root of unity:
\begin{displaymath}
\begin{array}{cccc}
e_1(\quad)e_1 & e_1(\quad)e_2 & e_1(\quad)e_3 & e_1(\quad)i \\
e_2(\quad)e_1 & e_2(\quad)e_2 & e_2(\quad)e_3 & e_2(\quad)i \\
e_3(\quad)e_1 & e_3(\quad)e_2 & e_3(\quad)e_3 & e_3(\quad)i \\
i(\quad)e_1 & i(\quad)e_2 & i(\quad)e_3 & i(\quad)i 
\end{array}
\end{displaymath}

It is easy to calculate explicitly these matrices. The whole set of the 
sixteen Conway matrices can be found in, e.g., Ref. \cite{JS72}. We only give 
below the four matrices used in Sec. \ref{s:dieq}
\begin{displaymath}
 \begin{array}{cc}
 e_3(\quad)e_2=\left(
          \begin{array}{cccc}
           0 & 0 & 0 & -1 \\
           0 & 0 & -1 & 0 \\
           0 & -1 & 0 & 0 \\
           -1 & 0 & 0 & 0 
           \end{array}
           \right)  
& 
 \; e_1(\quad)i=\left(
           \begin{array}{cccc}
           0 & \ 0 & 0 & \ i \\
           0 & \ 0 & -i & \ 0 \\
           0 & \ i & 0 & \ 0 \\
           -i & \ 0 & 0 & \ 0
           \end{array}           
           \right)  
\end{array}
\end{displaymath}

\begin{displaymath}
 \begin{array}{cc}
\; \;  e_3(\quad)e_1=\left(
           \begin{array}{cccc}
           0 & \ 0 & -1 & \ 0 \\
           0 & \ 0 & 0 & \ 1 \\
           -1 & \ 0 & 0 & \ 0 \\
           0 & \ 1 & 0 & \ 0
           \end{array}
           \right) 
& 
\, e_3(\quad)e_3=\left(
           \begin{array}{cccc}
           \ 1 & \ 0 & 0 & 0 \\
           \ 0 & \ 1 & 0 & 0 \\
           \ 0 & \ 0 & -1 & 0 \\
           \ 0 & \ 0 & 0 & -1
           \end{array}
           \right) 
 \end{array}
\end{displaymath}

\subsection{Quaternionic derivation}
\label{ss:quan}

The careful reader will have noticed that, in the whole article, we 
have avoided to define functions of a quaternionic variable which 
we should have been led to derive with respect to this variable. The 
reason is simply that, owing to the non-commutativity of the quaternionic 
product, the functions of a quaternionic variable which possess 
derivatives, i.e., for which the right and left derivatives are equal, 
are only the constant and linear functions. In fact, this is a general 
property shared by every number system with a non-commutative product 
\cite{GS93}. \\

However, the bi-quaternionic wave-function $\psi$, which is defined 
as a field in the four-dimensional space-time, i.e., of which the 
quaternionic components, ($\psi^{\nu}_0, \psi^{\nu}_1, \psi^{\nu}_2, 
\psi^{\nu}_3$) of each space-time component $\psi^{\nu}$, are complex 
functions of the coordinates $X^{\mu}$, can be derived with respect to these 
coordinates. For instance, $\partial_{\mu}\psi^{\nu}$ writes
\begin{eqnarray}
\partial_{\mu}\psi^{\nu}&=& {\partial  \psi^{\nu}_0 \over {\partial X^{\mu}}}
(X^0,X^1,X^2,X^3) + e_1 {\partial \psi^{\nu}_1\over {\partial X^{\mu}}}
(X^0,X^1,X^2,X^3) \nonumber \\
&+& e_2 {\partial \psi^{\nu}_2\over {\partial X^{\mu}}}(X^0,X^1,X^2,X^3) + 
e_3 {\partial \psi^{\nu}_3\over {\partial X^{\mu}}}(X^0,X^1,X^2,X^3).
\label{eq.121}
\end{eqnarray}

We are therefore allowed to apply to the components of this function all the 
ordinary complex derivation rules with respect to the coordinates. \\

On the contrary, any relation implying an application to a quaternionic 
function, other than the constant and linear functions, of ordinary 
derivation rules with respect to the quaternionic variable itself (see, e.g., 
Eq.~(\ref{eq.47}), which is however perfectly accurate for a complex $\psi$)  
would be uncorrect. It is easy to verify that our calculations leading to the 
quaternionic Klein-Gordon and Dirac equations are all along in perfect 
agreement with this constraint.

\end{document}